\documentclass[12pt]{article}
\usepackage{cite}

\usepackage{amsmath, amsthm, amssymb,slashed,hyperref}
\usepackage{ifpdf}
\ifpdf
  \usepackage[pdftex]{graphicx}
  \usepackage{epstopdf}
\else
  \usepackage[dvips]{graphicx}
\fi
\parskip=\baselineskip
\def\Bbb{\mathbb}

\def\16{{\bf 16}}
\def\1{{\bf 1}}
\def\2{{\bf 2}}
\def\3{{\bf 3}}
\def\4{{\bf 4}}
\def\MM{{\mathfrak M}}
\def\M{\MM}

\def\tilde{\widetilde}

\font\teneurm=eurm10 \font\seveneurm=eurm7 \font\fiveeurm=eurm5
\newfam\eurmfam
\textfont\eurmfam=\teneurm \scriptfont\eurmfam=\seveneurm
\scriptscriptfont\eurmfam=\fiveeurm

\font\teneusm=eusm10 \font\seveneusm=eusm7 \font\fiveeusm=eusm5
\newfam\eusmfam
\textfont\eusmfam=\teneusm \scriptfont\eusmfam=\seveneusm
\scriptscriptfont\eusmfam=\fiveeusm
\def\eusm#1{{\fam\eusmfam\relax#1}}
\font\tencmmib=cmmib10 \skewchar\tencmmib='177
\font\sevencmmib=cmmib7 \skewchar\sevencmmib='177
\font\fivecmmib=cmmib5 \skewchar\fivecmmib='177
\newfam\cmmibfam
\textfont\cmmibfam=\tencmmib \scriptfont\cmmibfam=\sevencmmib
\scriptscriptfont\cmmibfam=\fivecmmib

\numberwithin{equation}{section}

\def\veps{\varepsilon}
\def\U{{{\eusm U}}}
\def\VV{{\mathcal V}}
\def\WW{{\mathcal W}}
\def\t{\widetilde}
\def\S{{\mathcal S}}

\def\PP{{\mathcal P}}

\def\vp{\varphi}

\begin{document}
\begin{titlepage}
\begin{flushright}
hep-th/yymm.nnnn
\end{flushright}
\vskip 1.5in
\begin{center}
{\bf\Large{Filling The Gaps With PCO's}}
\vskip0.5cm 
{Ashoke Sen$^1$ and Edward Witten$^2$} 
\vskip.5cm
 {\small{\textit{$^1$Harish-Chandra Research Institute, Chhatnag Road, Jhusi, Allahabad 211019, Indis}}}
 \vskip.2cm
 {\small{\textit{$^2$School of Natural Sciences, Institute for Advanced Study, Princeton NJ USA 08540 \\}}
}
\end{center}
\vskip.5cm
\baselineskip 16pt
\begin{abstract}
Superstring perturbation theory is traditionally carried out by using picture-changing operators (PCO's) to integrate over odd moduli.
Naively the PCO's can be inserted anywhere on a string worldsheet, but actually a constraint must be placed on
PCO insertions to avoid spurious singularities.  Accordingly, it has been long known  that the simplest version of the PCO procedure
is valid only locally on the moduli space of Riemann surfaces, and that a correct PCO-based algorithm to compute scattering amplitudes
must be based on piecing together local descriptions.  Recently, ``vertical integration'' was proposed as a relatively simple method to do this.
Here, we spell out in detail what vertical integration means if carried out systematically.  This involves a hierarchical procedure with corrections of high
order.  One might anticipate such a structure from the viewpoint of super Riemann surfaces.
\end{abstract}
\end{titlepage}

\def\RRR{{\Bbb R}}
\newcommand{\eps}{\epsilon}
\newcommand{\vareps}{\varepsilon}

\newcommand{\UU}{{\mathcal U}}
\newcommand{\XX}{{\mathcal X}}
\newcommand{\OO}{{\mathcal O}}
\newcommand{\YY}{{\mathcal Y}}
\newcommand{\AAA}{{\mathcal A}}

\newcommand{\wt}{\widetilde}
\newcommand{\wh}{\widehat}

\newcommand{\be}{\begin{equation}}
\newcommand{\ee}{\end{equation}}
\newcommand{\ben}{\begin{eqnarray}\displaystyle}
\newcommand{\een}{\end{eqnarray}}

\newcommand{\refb}[1]{(\ref{#1})}
\newcommand{\p}{\partial}
\newcommand{\sectiono}[1]{\section{#1}\setcounter{equation}{0}}
\newcommand{\subsectiono}[1]{\subsection{#1}\setcounter{equation}{0}}

\tableofcontents   

\section{Introduction}\label{intro}

Superstring perturbation theory is traditionally constructed in an elegant framework of superconformal field theory,
with insertions of picture-changing operators (PCO's) as well as vertex operators for physical states \cite{FMS}.  The PCO's give
a method of integration over the odd moduli of a super Riemann surface \cite{VV}.  

Naively, the PCO's can be inserted at arbitrary
positions on a superstring worldsheet, but it has been known since the 1980's that this is oversimplified.  The measure on the moduli
space of Riemann surfaces that is constructed using PCO's has spurious singularities if two PCO's collide, and also if a certain global
condition is obeyed.\footnote{The global condition that leads to a spurious singularity  says that the superconformal ghost field $\gamma$ has a zero-mode if it is allowed to have simple
poles at the positions of PCO's.  But that fact is really not important for the present paper.}  For this paper, it suffices to know
that the locus of spurious singularities is of  complex codimension 1 or real codimension 2
and has a reasonable behavior at infinity on moduli space.\footnote{The simplest way to explain what one means by ``reasonable behavior'' is to say
that the bad set is an orbifold of complex codimension 1 or 
real codimension 2 even if one compactifies the moduli space of Riemann surfaces by allowing the usual degenerations.}   Actually,
the locus of spurious singularities is rather complicated and appears to have few useful properties beyond what we have just stated.  

To get a correct, gauge-invariant method of computing superstring scattering amplitudes, it is desirable
to avoid spurious singularities.   For topological reasons, a choice of PCO locations that avoids spurious singularities exists only locally on moduli space.
Accordingly, it has been understood since the 1980's that a correct method of computation based on PCO's has to be based on piecing together local
descriptions.  

A relatively simple method to piece together the local descriptions was proposed recently \cite{Sen} in the form of ``vertical integration.'' However, only
the basic idea of vertical integration was described.  Here, we explain systematically what vertical integration means if carried out in full.  An
inductive procedure is involved with corrections, in a certain sense, of all orders (bounded by the number of PCO's).
The need for corrections of  high order may come as a surprise to some readers.  However, this should be anticipated based on what was understood
in the old literature, and is fairly clear from the point of view of super Riemann surfaces.

In section \ref{goal}, we recall the basic idea of vertical integration. 
In section \ref{amplitude}, we describe the procedure systematically to all orders.
The construction described in section \ref{amplitude} requires making some 
choices for the ``vertical segment,''
and in section \ref{sequiv} we show that the 
scattering amplitude is independent of these choices. 
The measure that the procedure of section \ref{amplitude}  
generates on the moduli space of ordinary Riemann
surfaces is discontinuous, and this is compensated by additional terms that take the
form of integrals over subspaces of the moduli space of  codimension $\geq 1$.
In section \ref{smooth}, we describe a generalization of this procedure that generates 
a smooth measure on the moduli space and show that the procedure described in section 
\ref{amplitude} can be regarded as a special case of this. 
In section \ref{decoupling}, we show gauge invariance of the amplitude defined in
section \ref{smooth}.
In section \ref{super}, we explain why the
inductive or hierarchical procedure that we follow would be expected from the point of view of super Riemann surfaces.  

In this paper, we ignore the fact that the moduli space $M$ of Riemann surfaces is not compact.  This noncompactness arises from the fact
that the string worldsheet $\Sigma$ can degenerate, and is associated to the infrared behavior of string theory.  This infrared behavior
has been much analyzed in the literature and will not be considered here.  We simply remark that everything we say must be supplemented
with some fairly well-known conditions on the behavior of PCO's in the limit that $\Sigma$ degenerates.  

A  hierarchy of corrections somewhat similar  to what we describe here was used in \cite{EKS} to construct
a field theory of the NS sector of superstring theory.  Each string field theory diagram
parametrizes in a relatively simple way a piece of the moduli space of bosonic Riemann surfaces
and comes with a relatively natural choice of PCO insertions suitable for that piece.  On the boundaries
of the parts of moduli space parametrized by different diagrams,
the PCO choices do not fit together properly.  In \cite{EKS}, a hierarchy of corrections was introduced to compensate for this.

\section{Overview} \label{goal}

Before describing vertical integration in its most 
general form, we shall discuss some simple cases 
explicitly and explain
the issues one faces in extending to 
more general cases. 
Let us denote by $\XX(z)$ the PCO 
inserted at the point $z$ in a string worldsheet $\Sigma$. 
We can express this as
\be \label{eqxi}
\XX(z) = \{ Q_B, \xi(z)\}\, ,
\ee
where $\xi(z)$ is a fermion field of dimension (0,0) that arises from the bosonization
of the superghost system. $\xi(z)$ is an operator defined in the large Hilbert space of Friedan, Martinec, and Shenker \cite{FMS}.
We shall work in the small Hilbert space\footnote{Only the small Hilbert space appears to have a natural interpretation in terms of
super Riemann surfaces, so from that point of view one expects that all important formulas can be written in terms of operators of the small
Hilbert space.},  where one removes the zero-mode of $\xi$
from the spectrum of operators, so that only the derivatives of $\xi$ are valid operators.
All our analysis will involve only such operators. However, we shall make use of
the fact that the periods of the closed 1-form  $\partial\xi$  
vanish on any Riemann surface, even in the presence of
punctures labeled by operators of the small Hilbert space. Thus operators of the form $\xi(u) -\xi(v)\equiv
\int_v^u \p\xi(z) dz$ are well defined in the small Hilbert space without having to 
specify the contour of integration from $u$ to $v$.

Now consider a situation where the moduli space $M$ over which we integrate has real
dimension $n$
and suppose further that the correlation function of interest requires
insertion of only one PCO.  
 Each point $m\in M$ determines a Riemann surface $\Sigma(m)$, and the one PCO that we need can be inserted
 at an arbitrary point  $z\in\Sigma(m)$ except that we must avoid a bad set of (real) codimension 2 at which there are spurious
 singularities.  As $\Sigma(m)$ has dimension 2,
 the bad set consists of finitely many points in each $\Sigma(m)$.
 
  We denote by $Y^{}$ a fiber bundle with base $M$ and 
fiber $\Sigma(m)$:
\be\label{fiiberb}\begin{matrix}\Sigma(m)&{ \longrightarrow}& Y\cr && ~~\Big{\downarrow}\vp \cr && M.\end{matrix}\ee
We also denote as $X$ the subspace of $Y$ in which, in each fiber, one deletes the bad points at which the PCO
should {\it not} be inserted.  

We denote local coordinates on $X$ as $(m;a)$, with $m\in M$ and $a\in\Sigma(m)$.
$X$ is not a fiber bundle over $M$, because as one varies $m\in M$, the bad points in $\Sigma(m)$
can collide.  However, there certainly is a map $\vp:X\to M$. This is the map that forgets where the PCO
is inserted; in local coordinates, it maps $(m;a)$ to $m$.

Suppose that $M$ is of real dimension $n$. The path integral with one PCO insertion at $a\in M(m)$ 
(and all external vertex operators on-shell)
naturally computes for us  a closed  $n$-form on $X$:
\be \label{e1.2}
\omega_n(m;a) \equiv 
\left\langle (\XX(a) - \p\xi(a) da) \wedge \OO\right\rangle_n \, .
\ee
Here $ \langle ~~\rangle$ denotes a CFT correlation function on $\Sigma(m)$;
$\OO$ is a formal sum of operator-valued $k$-forms on $M$  
for all $k$ between 0 and $n$,
constructed from 
insertions of $b$-ghosts and possible 
on-shell vertex operators for external states.  
The subscript $n$ denotes that we have to extract the $n$-form part of this 
expression.\footnote{In fact, the $k$-form parts of this expression for others values of $k$, 
which we may call $\omega_k(m;a)$, are also useful {\it e.g.} in the proof of decoupling of
pure gauge states.  This is because $\omega_k$ satisfies the useful relation 
$\omega_k(Q_B|\Phi\rangle)= (-1)^k d\omega_{k-1}(|\Phi\rangle)$. Here $|\Phi\rangle$
denotes the collection of all external states and $Q_B$ is the total BRST operator
acting on all the external states. \label{fo4}}
The precise form of $\OO$ and the procedure to extract the closed $n$-form $\omega_n(m;a)$
is well-known and will not be described here.   

The subtlety of superstring perturbation theory in the PCO formalism arises because the PCO formalism naturally
constructs a closed $n$-form on $X$, not on $M$.  Ideally one would want  an $n$-form on $M$, which would automatically be closed
for dimensional reasons, and which would be integrated over $M$ to compute a scattering amplitude.  

How can we eliminate the dependence on $a$?  If we had a section $s:M\to X$
of the map $\vp:X\to M$, which concretely would be given in local coordinates by a formula\footnote{$M$ and $X$ are complex manifolds, 
but the section $s$ (or  equivalently the function $s(m)$) is not assumed to be holomorphic.} $a=s(m)$, then we could pull back
$\omega_n(m;a)$ to an $n$-form on $M$ and define the scattering amplitude as
\be\label{defint}\int_M s^*(\omega_n)=\int_M\omega_n(m,s(m)). \ee  
Since
$\omega_n$ is closed, this definition of the scattering amplitude is invariant under small changes in $s$. 
(From this point of view,
if there are topologically distinct choices of $s$ 
they might lead to different but equally well-defined results for the scattering amplitude.)  Moreover,
$\omega_n$ and therefore $s^*(\omega_n)$ changes by an exact form if one makes gauge transformations for some of the external vertex
operators, so the scattering amplitude defined this way would be  gauge-invariant.

\begin{figure}
\vskip -1in
 \begin{center}
   \includegraphics[width=3in]{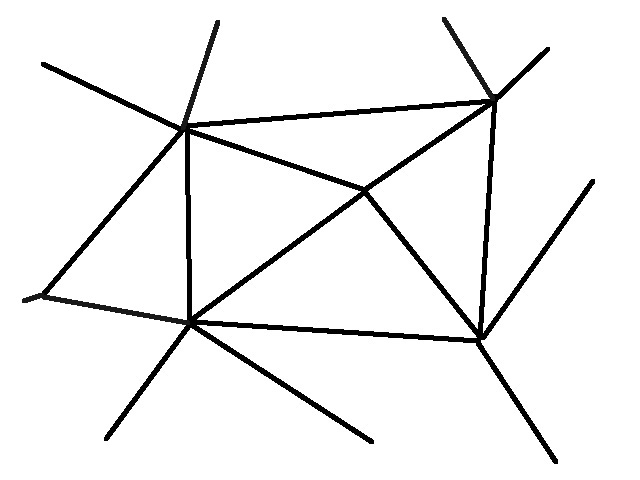}
 \end{center}
\caption{\small  A triangulation of a two-dimensional surface. }
 \label{triangles}
\end{figure}

 In general, the map $\vp$ does not
have a global section, but if we choose a sufficiently fine triangulation of $M$ (fig. \ref{triangles}) then
on each triangle, there will be a local section. This is just the statement that
on a sufficiently small triangle,  we can choose the PCO location as a continuous
function of $m$ while avoiding the bad points.\footnote{The term ``triangle'' assumes that $M$ has dimension $n=2$.  The $n$-dimensional generalization
of a triangle is called a simplex. In the present introductory explanation, we use two-dimensional terminology.}

Let $T_1 $ be one such triangle with local section $a=s^1(m)$.  The contribution to an on-shell 
amplitude from the triangle $T_1 $ with the PCO insertion at $a=s^1(m)$ can be expressed as
\be \label{e1.1}
\int_{T_1}  \omega_n(m;s^1(m)).
\ee

Now suppose that $T_2$ is a  second triangle  which shares a common boundary
$B$ with $T_1 $, and let $s^2$ denote a local section on
$T_2$. Then the contribution to the amplitude from $T_2$, computed with this local section, will be given by
\be 
\label{e1.3}
\int_{T_2} \omega_n(m;s^2(m))\, . 
\ee
Since $s^1(m)$ and $s^2(m)$ do not in general agree on the boundary $B$, the full amplitude must be
obtained by summing over contributions from different triangles together with appropriate
correction factors from the boundaries between the triangles.
 
Vertical integration is a prescription for determining these corrections. 
We  ``fill the gap'' in the integration cycle on $Y^{}$ by
drawing a vertical segment $U$.
$U$ is constructed by 
connecting the point $s^1(m)\in \Sigma(m)$ to 
$s^2(m)\in \Sigma(m)$
by a curve $C(m)\in \Sigma(m)$ for each\footnote{If the triangles $T_1$ and $T_2$ and therefore the boundary $B$ are small
enough, there is no problem in making $C(m)$ vary smoothly with $m$.  But in a moment we will see that this is not necessary.}
 $m\in B$, keeping away from the spurious singularities, and taking the collection
of all such curves: $\{C(m): m\in B\}$.  We parametrize $U$ by $m\in B$ and a variable $u\in [0,1]$ that labels the position
along the curve $C(m)$.  
The correction term associated with the boundary
$B$ is now taken to be given  by the integral of $\omega_n(m;a(u))$ over $U$.  
Using \refb{e1.2}, the integration over $u$ for fixed $m\in B$ can be
performed first, yielding the result
\be \label{e1.4}
\int_U \omega_n(m;u) = \int_B \, \langle (\xi(s^1(m)) - \xi(s^2(m))) \, \OO\rangle_{n-1}
\ee   The subscript just means that $\langle (\xi(s^1(m)) - \xi(s^2(m))) \, \OO\rangle_{n-1}$ is naturally an $(n-1)$-form.
Importantly, the right hand side does not depend on the choice of the paths $C(m)$, so we do not really need to pick a specific vertical segment $U$.

In general, $M$ may be triangulated with many triangles $T_i$, meeting in common boundaries $B_{ij}=T_i\cap T_j$ (most of the $B_{ij}$ are
empty). 
The full scattering amplitude is defined to be
\be\label{fuller}\sum_i\int_{T_i} \omega_n(m;s^i(m))+\int_{B_{ij} }\langle 
(\xi(s^i(m)) - \xi(s^j(m))) \, \OO\rangle_{n-1}.\ee 
Fixing the relative sign between the two terms requires fixing the orientation of
$B_{ij}$; this will be done carefully in section \ref{amplitude}. 
Standard arguments show that this formula is invariant under continuous changes of the $T_i$ and the $s^i$, and also is invariant under
gauge transformations of external state.

\begin{figure}
 \begin{center}
   \includegraphics[width=3.5in]{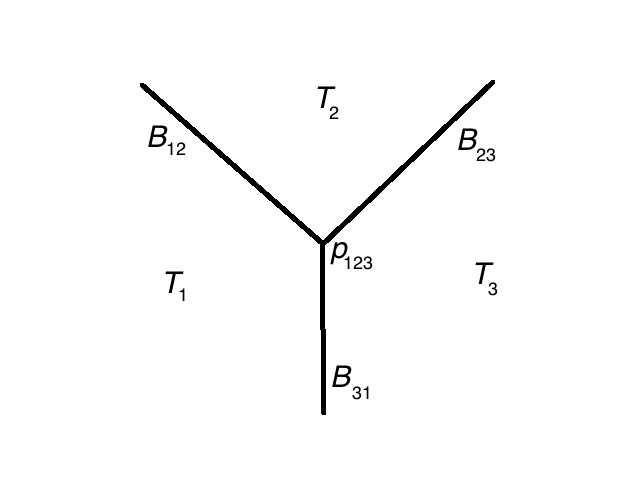}
 \end{center}\vskip-.5in
\caption{\small  A triple intersection of triangles. }
 \label{Triple}
\end{figure}

The logic behind this definition is as follows.  Over each double intersection $B_{ij}=T_i\cap T_j$ of triangles, we can define a ``vertical segment''
$U_{ij}$ as a union of paths from $s^i$ to $s^j$.  Now let us consider a  triple intersection $T_i\cap T_j\cap T_k$, with precisely three triangles
meeting at a common vertex (fig. \ref{Triple}).   This means that $B_{ij}$, $B_{jk}$, and $B_{ki}$ share a common endpoint $p_{ijk}$. (The case of
more than three triangles meeting at a vertex can be treated similarly.) It then makes
sense to ask if $U_{ij}$, $U_{jk}$, and $U_{ki}$ agree at $p_{ijk}$,
i.e.\ for each $m\in p_{ijk}$ the paths from $s_i(m)$ to $s_j(m)$ 
in $U_{ij}$, $s_j(m)$ to $s_k(m)$
in $U_{jk}$ and $s_k(m)$ to $s_i(m)$
in $U_{ki}$ together describe zero path.  If they do (for all triples $ijk$), then the triangles $s^i(T_i)$ and the
vertical segments $U_{ij}$ could be glued together to make a closed cycle $S\subset X$.  One would then define the scattering
amplitude as $\int_S\omega_n$.  This actually would agree with eqn. (\ref{e1.3}), since under the stated assumptions, $S$ could be slightly
perturbed to be a section $s:M\to X$.  It would clearly also agree with eqn. (\ref{fuller}), which expresses the scattering amplitude
as an integral over $S=\cup_i T_i\cup_{jk}U_{jk}$.  
 In reality, it may
 not possible to make the $U_{ij}$'s agree at triple intersections 
 since there may be
 a topological obstruction to finding a global
 section $s$, but because the formula of eqn. (\ref{fuller}) does not depend on the
 choices of the $U_{ij}$, this version of the formula makes sense anyway and has the same properties as if the $U_{ij}$ did agree on triple intersections. 
 In fact, we can study each triple intersection independently of the others, and at any one triple intersection, one can arrange so that the $U_{ij}$ do agree.

As long as only one PCO is needed, this is the end of the story.  The situation gets more complicated when there are more
PCO's. First of all, the generalization of
\refb{e1.4} now is ambiguous since $s^1(m)$ and $s^2(m)$ each will represent a collection of PCO's, and the integral in \refb{e1.4} depends
on the order in which we move the PCO's. 
Second, we may need additional correction terms from codimension $\ge 2$ subspaces
where three or more triangles meet. We can illustrate both these issues 
by considering the case
where we
need two PCO insertions in the correlator. 
In this case the role of $Y^{}$ is played by  the bundle whose base
is $M$ and whose fiber is $\Sigma(m)\times \Sigma(m)$.  As before, $X^{}$ is
obtained by excluding from $Y^{}$ certain codimension $ 2$ subspaces on which we
encounter spurious poles. 
The local coordinates of $X^{}$ can still be denoted as $(m;a)$ but now
$a$ stands for a pair of PCO
locations $(z_1, z_2)$. Similarly the choice of a local section $s^1$ on $T_1 $ will
now specify a pair of points $(z_1^{(1)}(m)\in \Sigma(m), z_2^{(1)}(m)\in \Sigma(m))$ 
avoiding spurious poles for 
each $m\in T_1 $, and
the choice of a local section $s^2$ on $T_2$ will
specify a pair of points $(z_1^{(2)}(m)\in \Sigma(m), z_2^{(2)}(m)\in \Sigma(m))$ 
avoiding spurious poles for $m\in T_2$.
The contribution to the amplitude from a given triangle $T_i $ is still given as 
\be\label{zold}\int_{T_i}(s^i)^*(\omega_n)=\int_{T_i}\omega_n(m;z_1^{(i)}(m),z_2^{(i)}(m)),\ee but $\omega_n(m;a)$ is now given by
\be \label{e1.10}
\omega_n(m;a) \equiv 
\left\langle (\XX(z_1) - \p\xi(z_1) dz_1)\wedge 
(\XX(z_2) - \p\xi(z_2) dz_2) \wedge \OO\right\rangle_n ,
\ee
with two PCO insertions.  
We can now try to determine the correction terms at the boundaries between
triangles by generalizing our prescription for vertical integration.
At an intersection of two triangles $T_1$ and $T_2$,  we again need to integrate over a ``vertical segment'' that fills in between
$s^1(T_1)$ and $s^2(T_2)$. For this, 
from each $m\in B$ we need to connect $(z_1^{(1)}(m), z_2^{(1)}(m))$ to 
$(z_1^{(2)}(m), z_2^{(2)}(m))$ by a path in
$\Sigma(m)\times \Sigma(m)$. 
But now, if we imitate the above procedure, the result will depend on the path.  It is easy to check, for example, that the
paths
\be \label{e123}
(z_1^{(1)}(m),z_2^{(1)}(m)) \to (z_1^{(2)}(m), z_2^{(1)}(m)) \to (z_1^{(2)}(m), z_2^{(2)}(m)) 
\ee
and
\be \label{e132}
(z_1^{(1)}(m),z_2^{(1)}(m)) \to (z_1^{(1)}(m), z_2^{(2)}(m)) \to (z_1^{(2)}(m), z_2^{(2)}(m)) 
\ee
give different results for the integral:
\ben \label{ecor1}
&& \int_B \bigg\langle
\left[(\xi(z_1^{(1)}) - \xi(z_1^{(2)})) (\XX(z_2^{(1)}) - \p\xi(z_2^{(1)}) dz_2^{(1)})
\right. \nonumber \\
&& \qquad \qquad \left. 
+ (\xi(z_2^{(1)}) - \xi(z_2^{(2)})) (\XX(z_1^{(2)}) - \p\xi(z_1^{(2)}) dz_1^{(2)})\right]  \wedge 
\OO\bigg\rangle_{n-1}
\een
and
\ben \label{ecor2}
&& \int_B \bigg\langle
\left[(\xi(z_2^{(1)}) - \xi(z_2^{(2)})) (\XX(z_1^{(1)}) - \p\xi(z_1^{(1)}) dz_1^{(1)})
\right.\nonumber \\
&& \qquad \qquad 
\left.+ (\xi(z_1^{(1)}) - \xi(z_1^{(2)})) (\XX(z_2^{(2)}) - \p\xi(z_2^{(2)}) dz_2^{(2)})\right]  
\wedge \OO \bigg\rangle_{n-1}
\een
where $dz_i^{(1)}$, $dz_i^{(2)}$ have to interpreted as their pullback to $B$.
Let us suppose that we have made some specific choice of the path for each boundary
separating a pair of triangles.
Now if we consider the subspace of $M$ where three
triangles $T_1 $, $T_2$ and $T_3$  meet, 
then the chosen path from $(z_1^{(1)}(m),z_2^{(1)}(m)) \to (z_1^{(2)}(m), z_2^{(2)}(m))$ 
on the boundary 
between $T_1 $ and $T_2$, together with the chosen path from
$(z_1^{(2)}(m),z_2^{(2)}(m)) \to (z_1^{(3)}(m), z_2^{(3)}(m))$ on the boundary between $T_2$ and $T_3$,
may not match the chosen path from 
$(z_1^{(1)}(m),z_2^{(1)}(m)) \to (z_1^{(3)}(m), z_2^{(3)}(m))$ on the
boundary between $T_1 $ and $T_3$. This means that when we regard the integrals as integrals 
over subspaces of $Y^{}$, then even after filling the gaps between the sections over
$T_1 $ and $T_2$, the sections over $T_2$ and $T_3$ and the sections over $T_1 $ and $T_3$, we 
are left with a gap over the common intersection of the three triangles. 
The earlier argument based on the path independence of \refb{e1.4} does not help us
since now the result does depend on some details of the path.
Thus we now need to
``fill this gap,'' leading to additional correction terms. 

In general, for computing an amplitude  with some given numbers of external legs of  Neveu-Schwarz or Ramond type
and given genus, we need a fixed number $K$ of PCO insertions.  
The analog of $Y$ in the above discussion is a fiber bundle over $M$ whose fiber is
 a  product 
 $\Xi(m)=\Sigma(m)\times \Sigma(m)\times\dots\times 
 \Sigma(m)$ 
 of $K$ copies of $\Sigma(m)$.   
 The analog of $X$ is obtained by omitting
 from each fiber of $\Xi(m)$ a codimension 2 subset on which spurious singularities arise.
We shall denote a point in $X$ by $(m;a)$ with $m\in M$, $a\in \Xi(m)$ and by
$\vp: X\to M$ the map that forgets $a$. 
In general, $\vp:X\to M$ 
does not have a global section but it has local
sections. Thus  if we triangulate $M$ (we follow a slightly different procedure in section \ref{amplitude}),
a local section will exist over each simplex (recall that a simplex is the $n$-dimensional analog of a triangle).
We can follow the procedure described above, integrating $(s^i)^*(\omega_n)$ over each simplex and making corrections
on the boundaries of simplices.  But now, further corrections will be needed
on higher
codimension subspaces where the boundaries meet.  In general, one needs corrections on codimension $k$ subspaces for all $k\le K$. 
The main goal of this paper to give a systematic procedure for constructing these
correction terms and to show that once all the corrections are added, the result has the
desired properties of the string amplitudes. In particular, it is gauge-invariant and free from any ambiguity.

\section{General Procedure} \label{amplitude}

In this section, we shall generalize the ideas of section \ref{goal}
 to arrive at
a complete prescription for computing the amplitude.

\subsection{Dual Triangulations} \label{triangulations}

\begin{figure}
 \begin{center}
   \includegraphics[width=3.5in]{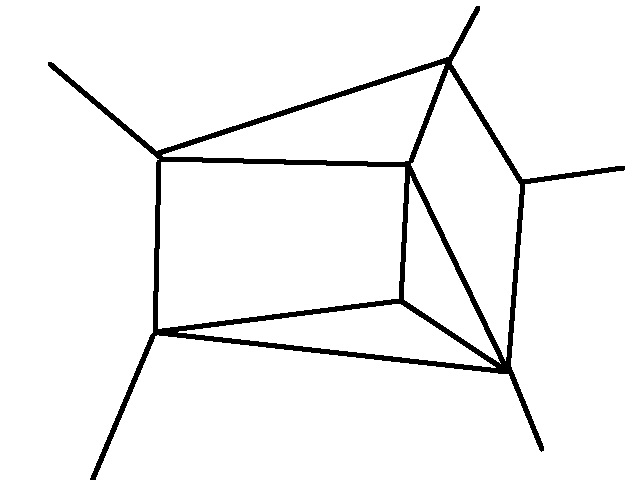}
 \end{center}
\caption{\small  A covering of a two-dimensional surface by more general polygons. }
 \label{poly}
\end{figure}

For carrying out this program,
roughly speaking, we will use a triangulation of $M$, but actually triangulation is not precisely the most convenient notion.
  To ``triangulate'' an $n$-manifold $M$ means to build it by gluing together
simplices, or simply by triangles if $n=2$.
 In a triangulation, any number of simplices might meet at a vertex.  Instead of triangles, we might
cover $M$ by more general polyhedra again in general with any number of building blocks meeting at a vertex.  This is sketched in 
two dimensions in fig. \ref{poly}.
The analog of this in dimension $n$ is to use $n$-dimensional polyhedra, perhaps of some restricted type, as the building blocks,
rather than $n$-simplices.

\begin{figure}
 \begin{center}
   \includegraphics[width=4.5in]{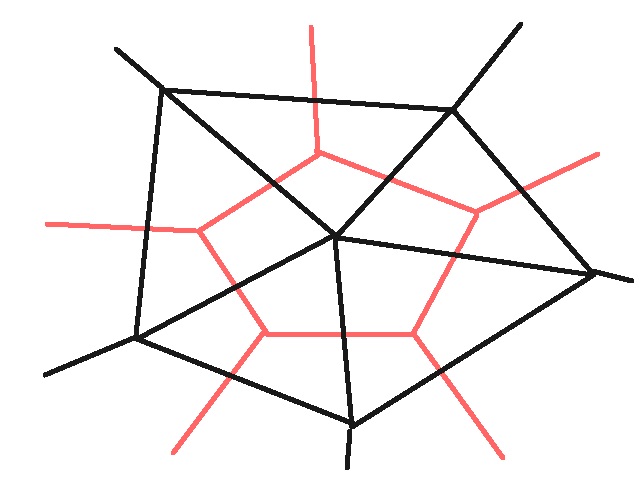}
 \end{center}
\caption{\small A triangulation (black) and the dual covering (red). We call the dual covering a dual triangulation.   Any vertex of one of the polygons making
up a dual triangulation is contained in precisely three of those polygons. }
 \label{braidbo2}
\end{figure}

For our purposes, we do not want an arbitrary covering by polyhedra, but the restriction to simplices is also not convenient.  
We use the fact that to a covering $\Lambda$ by polyhedra, we can associate a dual covering $\t\Lambda$.  In this duality, faces of dimension $k$ are replaced
by faces of dimension $n-k$ that meet them transversely.  In two dimensions, this means that a polygon in the covering $\Lambda$
 corresponds to a vertex in the dual
covering $\t\Lambda$, and vice-versa, while the edges $\Lambda$ meet the edges in $\t\Lambda$ transversely (fig.~\ref{braidbo2})).

\begin{figure}

 \begin{center}
   \includegraphics[width=3.5in]{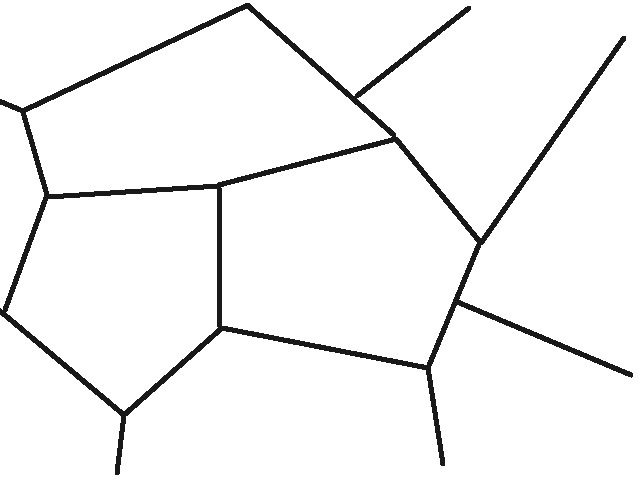}
 \end{center}
 \caption{\small  A dual triangulation of a two-dimensional surface is defined
by drawing a trivalent graph on the surface.  }
 \label{braidbo3}
\end{figure}

In the two-dimensional example shown in fig.~\ref{braidbo2}, the ``original'' covering $\Lambda$ is a triangulation.  This means generically that the dual covering 
$\t\Lambda$ is not a triangulation, but a covering by more general polygons.  However, $\t\Lambda$ has a useful
property: every vertex in $\t\Lambda$ is contained precisely in three polygons.  In two dimensions, a  covering of $M$ with this property can be
built by drawing a trivalent graph on $M$ (fig.~\ref{braidbo3})).

If $M$ is a manifold of any dimension $n$, by a ``dual triangulation,'' we mean a covering that is dual to a triangulation.
Thus, if $\Upsilon$ is a dual triangulation of $M$, then it is built by gluing together $n$-dimensional polyhedra along their 
boundary faces,
in such a way that for $k=1,\dots,n$, every codimension $k$ face of one of the polyhedra is contained in precisely $k+1$
polyhedra in $\Upsilon$.  This generalizes the fact that in dimension 
2, every edge in a dual triangulation is contained in two polygons and every vertex is contained in three polygons.  It will be technically easier
for us to use a dual triangulation rather than some other type of covering, because it is useful to have a bound on the number of polyhedra
that meet at a face of  given codimension.

\subsection{Basic Setup}

We pick a dual triangulation $\Upsilon$  of $M$ by gluing together polyhedra.   For $k=0,\dots ,n$, let $S_k$ be the set of codimension
$k$ faces of all the polyhedra that make up $\Upsilon$. So $S_0$ is the set of polyhedra, $S_1$ is the
set of their boundary faces, $S_2$ is the set of codimension 2 faces making up the boundaries of the faces in $S_1$, and so on. We denote a polyhedron
as $M_0^\alpha$, $\alpha\in S_0$, and denote by $M_k^{\alpha_0\cdots \alpha_k}$
the codimension $k$ face that is shared by the codimension zero faces
$M_0^{\alpha_0},\cdots, M_0^{\alpha_k}$.  We pick an orientation on $M$; this restricts to an orientation of
each  polyhedron $M_0^\alpha$, $\alpha\in S$.   We pick an orientation on 
$M_k^{\alpha_0\cdots \alpha_k}$ via the relation
\be \label{eorientation}
\p M_k^{\alpha_0\cdots \alpha_k} = 
-\sum_\beta M_{k+1}^{\alpha_0\cdots \alpha_k \beta} 
\ee
where the sum over $\beta$ runs over all codimension 0 faces $M_0^\beta$, distinct from $M_0^{\alpha_0},\dots, M_0^{\alpha_k}$, that have nonempty
intersection with  $M_k^{\alpha_0\cdots \alpha_k}$. These definitions imply that the
orientation of $M_k^{\alpha_0\cdots \alpha_k}$ changes sign
under $\alpha_i\leftrightarrow \alpha_j$
for any pair $(i,j)$.

We pick $\Upsilon$ to be fine enough so that the map $\varphi:X\to M$  
has a section  $s^\alpha$
over each of the polyhedra $M_0^\alpha$, $\alpha\in S_0$.  We need to impose further conditions on the $s^\alpha$.
To motivate the needed conditions, we consider the case of just two PCO's and examine the correction terms \refb{ecor1} or \refb{ecor2}
that are needed on the intersection $M_1^{12}=M_0^1\cap M_0^2$ of two polyhedra.
These
depend on two sets
of PCO data {\it e.g.} $(z_1^{(1)}(m), z_2^{(1)}(m))$ on $M_0^1$
and $(z_1^{(2)}(m), z_2^{(2)}(m))$ on $M_0^2$. 
The corrections involve mixed correlation 
functions involving products $\xi(z_1^{(2)}) \XX(z_2^{(1)})$ or $\xi(z_2^{(1)}) \XX(z_1^{(2)})$.
To ensure that
these are free from spurious singularities, it is not enough that $(z_1^{(1)}, z_2^{(1)})$ 
and $(z_1^{(2)},z_2^{(2)})$
separately describe configurations free from spurious singularities; we also require that
$(z_1^{(1)}, z_2^{(2)})$ and $(z_1^{(2)}, z_2^{(1)})$ describe configurations with the same property.\footnote{For this we use the fact that the locations of the spurious singularities
in the correlation functions involving products of $\xi$'s and $\XX$'s remain unchanged
if we replace $\xi$'s by $\XX$'s. This follows from the general form of the correlation
functions of these operators given in \cite{VV}.}

In order to describe the required condition for the general case, let us introduce
some notation. 
For some given $m\in M$, let $a^0,\cdots ,a^k$ denote $k+1$
possible PCO arrangements, with each $a^\alpha$ standing for a set of
$K$ points   $(z_1^{(\alpha)},\cdots 
z_k^{(\alpha)})$ with $z_i^{(\alpha)}(m)\in \Sigma(m)$. Now consider the 
$(k+1)^K$ possible arrangement of PCO's $(z_1,\cdots z_K)$ 
where each $z_i$ can take values
$z_i^{(\alpha_0)},\cdots z_i^{(\alpha_k)}$. We shall say that 
$(m;a^0,\cdots a^k)\in X^{(k+1)}$ if each of these $(k+1)^K$ 
PCO arrangements is
free
from spurious singularity.

We are now in a position to state the general condition on the sections $s^\alpha$
on the codimension zero faces. It states that
on a codimension $k$ face 
$M_k^{\alpha_0\cdots\alpha_k}$ 
that is shared by
$k+1$ codimension zero faces $M_0^{\alpha_0},\cdots M_0^{\alpha_k}$, 
the corresponding sections 
satisfy the restriction
\be \label{extracond}
(m; s^{\alpha_0}(m),\cdots s^{\alpha_k}(m))\in X^{(k+1)} \quad \hbox{for $m\in 
M_k^{\alpha_0\cdots\alpha_k}$}\, .
\ee
The existence of sections satisfying this condition (for a sufficiently fine dual triangulation $\Upsilon$) will be proved in section \ref{ssection}.

For  the condition \refb{extracond} to be meaningful,  we need to choose some ordering
of the PCO locations associated with each $s^\alpha$, since this condition is not invariant under 
permuting the PCO locations inside one $s^\alpha$
({\it e.g.} $z_1^{(\alpha)}\leftrightarrow z_2^{(\alpha)}$) 
keeping the other $s^\alpha$'s unchanged. We
shall assume that some specific ordering of the PCO's has been chosen on each 
codimension zero face $M_0^\alpha$
so that \refb{extracond} is meaningful.   However, the argument in section \ref{ssection} will
actually show that we can assume that the condition \refb{extracond} is satisfied independently for each possible permutation.

From the discussion involving eqns. \refb{e123} and \refb{e132}, we know  
that we need to choose additional data
to carry out the program of vertical integration.
This will be described next. 

\subsection{Additional Data}  \label{sdata}

Let $M_1^{\alpha\beta}$
be the codimension 1 face shared by the codimension 0 faces $M_0^\alpha$ and $M_0^\beta$.
Then on $M_1^{\alpha\beta}$ we need to choose a ``path'` $P_{\alpha\beta}$
from the PCO locations
 $(z_1^{(\alpha)},\cdots z_K^{(\alpha)})$ to $(z_1^{(\beta)}. \cdots z_K^{(\beta)})$. 
If we denote by $\Xi(m)$ the product $\Sigma(m)\times \dots\times \Sigma(m)$ of $K$ copies of $\Sigma(m)$,
 then $P_{\alpha\beta}$
can be regarded as a path in $\Xi$ from the PCO locations on $M_0^\alpha$
to the PCO locations on $M_0^\beta$. Once a path $P_{\alpha\beta}$ has been
chosen this way, we will choose $P_{\beta\alpha}$  be $-P_{\alpha\beta}$
i.e.\ the same path traversed in opposite direction.  The paths will be constructed by moving PCO's one at a time
from an initial location $z_j^{(\alpha)}$ (for some $j$) to a final location $z_j^{(\beta)}$.

It will be crucial that the construction depends only on the order in which the PCO's are moved between their
initial and final positions, and not on the precise path by which they are moved.   Even two
topologically distinct paths between $z_j^{(\alpha)}$ and $z_j^{(\beta)}$
 will be equivalent for our application.
This is due to the fact that expressions of the form \refb{e1.4} or \refb{ecor1}, \refb{ecor2} that result
from integration over a segment of the path in which just one PCO is moved depend only on the
initial and final PCO locations and not on the path connecting them. For this reason it will be useful
to develop a symbolic 
representation of these paths that only captures the relevant information without
any unnecessary data. This can be done as follows. To 
each codimension 1 face $M_1^{\alpha\beta}$ 
associate a $K$-dimensional Euclidean space $\RRR^K$, and represent a PCO configuration
$(z_1,\cdots z_K)$ with each $z_i$ taking values $z_i^{(\alpha)}$ or $z_i^{(\beta)}$ by an
integer lattice point in $\RRR^K$, with the $i$-th coordinate 
being 0 if $z_i=z_i^{(\alpha)}$ and
1 if $z_i=z_i^{(\beta)}$. Thus for example the origin represents the PCO configuration 
$(z_1^{(\alpha)}, z_2^{(\alpha)}, \cdots z_K^{(\alpha)})$ and the point $(1,1,\cdots 1)$ 
represents the PCO configurations $(z_1^{(\beta)}, z_2^{(\beta)}, \cdots z_K^{(\beta)})$. The
path $P_{\alpha\beta}$ now can be represented by 
a path $Q_{\alpha\beta}$ in $\RRR^K$ connecting
the origin to $(1,1,\cdots 1)$, lying along 
the edges of a unit hypercube.  
Given any such path $Q_{\alpha\beta}$, it captures all the relevant information about 
$P_{\alpha\beta}$ even though in actual practice there are many topologically distinct 
paths on $\Xi$ associated with a given $Q_{\alpha\beta}$. All of these paths will 
give the same  
result for the integral that will be written down in section \ref{integral}. 
We pick a particular $Q_{\alpha\beta}$ for each pair $\alpha,\beta$, with $Q_{\beta\alpha}=-Q_{\alpha\beta}$.

In eqs.\refb{e123} and 
\refb{e132}, we considered an example with  $K=2$. The path \refb{e123} 
will be represented as $(0,0)\to (1,0)\to (1,1)$ and the path \refb{e132} will be
represented as $(0,0)\to (0,1)\to (1,1)$.

To fully define vertical integration, we will need to refine this procedure and make some additional choices.
Consider a particular codimension 2 face
$M_2^{\alpha\beta\gamma}$. Having picked a section $s^\alpha$ over each $M_0^\alpha$,
we have on $M_2^{\alpha\beta\gamma}$ three sets of PCO data: $(z_1^{(\alpha)}, \cdots z_K^{(\alpha)})$, 
$(z_1^{(\beta)}, \cdots z_K^{(\beta)})$,  $(z_1^{(\gamma)}, \cdots z_K^{(\gamma)})$.
We now consider the $3^K$ PCO configurations $(z_1,\cdots z_K)$ with $z_i$ taking
values $z_i^{(\alpha)}$, $z_i^{(\beta)}$ or $z_i^{(\gamma)}$ for each $i$ and 
represent them as follows as points in $\RRR^K$: the $i$-th coordinate is assigned
value 0 if $z_i$ is $z_i^{(\alpha)}$, 1 if $z_i$ is $z_i^{(\beta)}$ and 2 if $z_i$ is 
$z_i^{(\gamma)}$. Thus in this description 
the path $P_{\alpha\beta}$ can be represented by a path 
$Q_{\alpha\beta}$ from
the origin $(0,0,\cdots,0)$ to the point $(1,1,\cdots, 1)$,
the path $P_{\beta\gamma}$ is represented by a path $Q_{\beta\gamma}$ from
$(1,1,\cdots ,1)$ to $(2,2,\cdots ,2)$ and $P_{\gamma\alpha}$ is represented by a path 
$Q_{\gamma\alpha}$
from
$(2,2,\cdots, 2)$ to $(0,0,\cdots ,0)$. Together they form a closed path in $\RRR^K$.
We now need to choose a  subspace $Q_{\alpha\beta\gamma}$ 
of $\RRR^K$, satisfying the following properties:
\begin{itemize}
\item The boundary of $Q_{\alpha\beta\gamma}$ is given by
\be \label{ebd}
\p Q_{\alpha\beta\gamma} = - Q_{\alpha\beta}-Q_{\beta\gamma}-Q_{\gamma\alpha}\, .
\ee
\item $Q_{\alpha\beta\gamma}$ is 
made of a collection of rectangles whose vertices are integer points 
of $\RRR^K$ with coordinates 0, 1 or 2 and whose sides lie 
along some coordinate axes, i.e. 
along each rectangle only two of the coordinates of $\RRR^K$ vary.
\item Once $Q_{\alpha\beta\gamma}$ has been chosen, we define 
$Q_{\beta\alpha\gamma}$ to be $-Q_{\alpha\beta\gamma}$. More generally
$Q_{\alpha\beta\gamma}$ is chosen to be antisymmetric under the
exchange of any pair of its subscripts.
\end{itemize}

For given $\alpha,\beta,\gamma$, it is possible to choose a $Q_{\alpha\beta\gamma}$ satisfying
these conditions essentially because the closed path $Q_{\alpha\beta}+Q_{\beta\gamma}+Q_{\gamma\alpha}$
is contained in a certain finite collection of unit squares (the squares in $\RRR^K$ whose corners have coordinates
0, 1, or 2), and this collection is simply-connected.
The choice of $Q_{\alpha\beta\gamma}$
is of course is not unique; there are many unions of
rectangles leading to the same result, just as there were many choices of $Q_{\alpha\beta}$.
Given a choice of $Q_{\alpha\beta\gamma}\in \RRR^K$, we can associate with it a two-dimensional
region $P_{\alpha\beta\gamma}$ of $\Xi$ composed of ``rectangular regions'' whose corners 
correspond to PCO locations $(z_1,\cdots z_K)$ with each $z_i$ taking values 
 $z_i^{(\alpha)}$, $z_i^{(\beta)}$ or $z_i^{(\gamma)}$, and along which only two of the $z_i$'s
 vary.

We continue in this way for higher codimensions.   Given  a codimension $k$ face 
$M_k^{\alpha_0\cdots \alpha_k}$ shared by $k+1$ codimension zero
faces $M_0^{\alpha_0},\cdots, M_0^{\alpha_{k}}$, we can represent the  PCO
locations determined by the sections $s^{\alpha_0},\dots, s^{\alpha_k}$ 
as integer points in $\RRR^K$, with the prescription that if 
the $i$-th PCO location is $z_i^{(\alpha_s)}$ then the $i$-th coordinate
is $s$. The analysis at the previous
step would have determined the $(k-1)$-dimensional subspaces 
$Q_{\alpha_0\cdots \alpha_{k-1}}$, $Q_{\alpha_0,\cdots
\alpha_{k-2}, \alpha_k}$ etc., each of which can be represented as 
$(k-1)$-dimensional subspaces of $\RRR^K$
composed of a union of hypercuboids\footnote{A hypercuboid is the multi-dimensional generalization of a rectangle.
In general, we will consider $k$-dimensional hypercuboids in $\RRR^n$ for various $k$.
Their corners will always lie in the lattice in  $\RRR^n$  consisting of points
with integer coefficients and their sides will be  parallel (or perpendicular) to each of the coordinate axes.
We describe this loosely by saying that the sides of the hypercuboid lie along coordinate axes.
Such hypercuboids are built by gluing together a certain number of adjacent, 
parallel unit $k$-dimensional
hypercubes in the lattice of integer points. 
 Because of this, one could express all statements in terms of unit hypercubes rather than hypercuboids.} with vertices given by integer points and in each hypercuboid only $k-1$
of the coordinates of $\RRR^K$ vary. We now have to choose a $k$-dimensional subspace 
$Q_{\alpha_0\cdots \alpha_k}$ of $\RRR^K$ satisfying the condition
\be \label{ebdfull}
\p Q_{\alpha_0\cdots \alpha_k} = -\sum_{i=0}^k (-1)^{k-i} 
Q_{\alpha_0\cdots \alpha_{i-1}\alpha_{i+1} \cdots \alpha_k}  
\, .
\ee
Furthermore we choose $Q_{\alpha_0\cdots \alpha_k}$ to be a union of $k$-dimensional hypercuboids 
with vertices at integer points and along each of which only $k$ of the coordinates of $\RRR^K$
vary. This can be mapped back to a $k$-dimensional subspace of $\Xi$ consisting of
``hypercuboid-shaped regions'' with vertices given by the PCO locations for which $z_i$ can take
one of the $k+1$ values $z_i^{(\alpha_0)},\cdots z_i^{(\alpha_k)}$ for each $i$ and along
each of these hypercuboid shaped regions only $k$ of the PCO locations vary. Finally, we choose
 $Q_{\alpha_0\cdots \alpha_k}$ to be antisymmetric under the exchange
of $\alpha_i$ and $\alpha_j$.

How far do we need to continue? First of all, it is clear that we must have $k\le n$
since $M_k$ has codimension $k$. But also we must have $k\le K$ since 
$Q_{\alpha_0\cdots\alpha_k}$ has dimension $k$. Typically in the situations
we encounter, we always have $K\le n$ and hence $k\le K$ is the bound we
need to satisfy.  That is why the examples of section \ref{goal} with $k=1,2$ did not
require developing the full procedure.

Once we have constructed the $Q_{\alpha_0\cdots \alpha_k}$'s, we can associate
with it a $k$-dimensional subspace $P_{\alpha_0\cdots\alpha_k}(m)$
of $\Xi(m)$ as follows. Since $Q_{\alpha_0\cdots\alpha_k}$ can be regarded as
a collection of hypercubes in $\RRR^K$ it is enough to prescribe how to construct
$k$-dimensional subspaces of $\Xi$ for each hypercube in $\RRR^K$
and then regard 
$P_{\alpha_0\cdots \alpha_k}$ as a union of these subspaces. 
For this we first replace $\Xi(m)$ by its universal cover $\t\Xi(m)$ 
by taking $K$ copies of the
universal cover $\t\Sigma(m)$ of $\Sigma(m)$, and represent each PCO location 
$z_i^{(\alpha)}$ for $1\le i\le K$, $0\le\alpha\le (k+1)$ by a point in $\t\Sigma(m)$.
This choice is not unique since each point in $\Sigma(m)$ has infinite number of
representatives on $\t\Sigma(m)$; 
we pick any one representative. This allows us to
represent the $(k+1)^K$ PCO arrangements 
$(z_1, \cdots z_K)$ -- with each $z_i$ taking values
$z_i^{(\alpha_0)},\cdots z_i^{(\alpha_k)}$ -- by $(k+1)^K$ points on $\t\Xi(m)$.
Now given a $k$-dimensional hypercube in $\RRR^K$ we first 
map its corner points to $\t\Xi$, taking the point $(\beta_1,\cdots \beta_K)\in\RRR^K$ to the point
$(z_1^{(\beta_1)},\cdots z_K^{(\beta_K)})$ on $\t\Xi(m)$. Next consider the dimension
one edges of the hypercube. Along each such edge, one of the coordinates of $\RRR^K$
vary. If the $i$-th coordinate varies then we map it to a curve in $\t\Xi(m)$ along which
only $z_i$ varies, keeping all $z_j$'s with $j\ne i$ constant. The end points of the
curve are fixed by the locations of the vertices but 
the shape of the curve in the $z_i$
plane can be chosen arbitrarily. After mapping all the dimension one edges to $\Xi$ 
this way, we turn to the dimension two faces. Along each face of the hypercube in
$\RRR^K$ only two of the coordinates vary. Suppose that the $i$-th and the $j$-th
coordinates vary along a particular face. We map it to a two dimensional subspace of
$\t\Xi$ along which only $z_i$ and $z_j$ vary leaving all other $z_k$'s fixed. The 
boundary of the two dimensional subspace is fixed by the choice of the dimension
one edges at the previous step, but how $z_i$ and $z_j$ vary in the interior can be
chosen arbitrarily. The maps for higher dimensional faces proceed in a similar
manner. For a dimension $\ell$ face of the hypercube in $\RRR^K$, along
which the $i_1,\cdots i_\ell$'th coordinates vary keeping the other coordinates
fixed, we associate an $\ell$-dimensional  
subspace of $\t\Xi$ along which $z_{i_1},\cdots z_{i_\ell}$
vary leaving the other coordinates fixed. The boundary of this subspace is fixed
by the choice made at the previous step, but the choice of how
$z_{i_1},\cdots z_{i_\ell}$ vary in the interior can be made arbitrarily. Proceeding
this way all the way upto $\ell=k$ 
we can construct the map of the entire $k$-dimensional hypercube
to a $k$-dimensional subspace of $\t\Xi$. After we have repeated this construction for
every hypercube contained in $Q_{\alpha_0\cdots \alpha_k}$, we can construct 
the $k$-dimensional subspace of $\t\Xi$ obtained by union
of these subspaces of $\t\Xi$. 
This can now be interpreted as a $k$-dimensional subspace of $\Xi$.
We call this $P_{\alpha_0\cdots \alpha_k}$.

As a consequence of
\refb{ebdfull}, the $P_{\alpha_0\cdots\alpha_k}$'s
constructed this way satisfy the identity:
\be \label{ebdP}
\p P_{\alpha_0\cdots \alpha_k} \simeq -\sum_{i=0}^k (-1)^{k-i} 
P_{\alpha_0\cdots \alpha_{i-1}\alpha_{i+1}\cdots \alpha_k}  
\, ,
\ee
where $\simeq$ symbol in \refb{ebdP} means that the boundary of 
$P_{\alpha_0\cdots \alpha_k}$ can be regarded as a collection of $(k-1)$-dimensional
subspaces of $\Xi$ whose corner points agree with the those of the right hand side
of \refb{ebdP}. 
However the hypercubes themselves may not be identical
since we might have used different choices for  constructing the faces of various
dimensions from the given corner points and 
might even have used different representatives for some of the PCO locations
on the universal cover of $\Sigma(m)$.
For example
$-P_{\alpha\beta}-P_{\beta\gamma}-P_{\gamma\alpha}$ constructed using this procedure
may even 
describe a non-contractible cycle of $\Xi$ in which case there is no subspace of
$\Xi$ whose boundary is given by this combination. However by choosing to define
$P_{\alpha\beta}$, $P_{\beta\gamma}$ and $P_{\gamma\alpha}$ 
on $M_2^{\alpha\beta\gamma}$ using different paths with the same
end-points we can make $-P_{\alpha\beta}-P_{\beta\gamma}-P_{\gamma\alpha}$
contractible and form the boundary of $P_{\alpha\beta\gamma}$.

Once we have chosen all the $Q_{\alpha_0\cdots \alpha_k}$ (and hence also the
$P_{\alpha_0\cdots \alpha_k}$) via this procedure, we can formally construct a
continuous integration cycle in $Y$ as follows.
First, for each codimension zero face $M_0^\alpha$, 
the section $s^\alpha$ gives a subspace of $Y$. Let us call this $\Sigma_\alpha$. 
In a generic
situation, $s^\alpha$ and $s^\beta$ 
will not match at the boundary $M_1^{\alpha\beta}$
separating $M_0^\alpha$ and $M_0^\beta$, leaving a gap in the integration cycle
between $\Sigma_\alpha$ and $\Sigma_\beta$.
We fill these gaps by including, for each $M_1^{\alpha\beta}$, a subspace 
$\Sigma_{\alpha\beta}$ of $Y$
obtained by fibering $P_{\alpha\beta}$ on $M_1^{\alpha\beta}$. However since on the
codimension 2 face $M_2^{\alpha\beta\gamma}$ 
$P_{\alpha\beta}$, $P_{\beta\gamma}$ and $P_{\gamma\alpha}$ enclose a
non-zero  subspace of $\Xi$,
the subspaces $\Sigma_{\alpha\beta}$,
$\Sigma_{\beta\gamma}$ and $\Sigma_{\gamma\alpha}$ will not meet. This gap
will have to be filled by the space $\Sigma_{\alpha\beta\gamma}$ obtained by
fibering $P_{\alpha\beta\gamma}$ over $M_2^{\alpha\beta\gamma}$. Proceeding
this way we include all subspaces $\Sigma_{\alpha_0\cdots\alpha_k}$ obtained by
fibering $P_{\alpha_0\cdots \alpha_k}$ on $M_k^{\alpha_0\cdots\alpha_k}$. This formally
produces a continuous integration cycle in 
$Y$.\footnote{This construction of a cycle is only 
formal since {\it e.g.} the $P_{\alpha\beta}$ that 
forms part of the boundary of $P_{\alpha\beta\gamma}$ may differ from the
$P_{\alpha\beta}$ that was fibered over $M_{\alpha\beta}$ to construct 
$\Sigma_{\alpha\beta}$ 
by non-trivial cycles on $\Xi$.}

We shall call the segments $\Sigma_{\alpha_0\cdots\alpha_k}$ for $k\ge 1$ vertical
segments. Typically these segments
pass through spurious poles  
and hence it may not be immediately obvious that this procedure can lead to a sensible
definition of a scattering amplitude.  However, we shall now show that
this can be done by generalizing what has been explained
in section \ref{goal}.

\subsection{Contributions from Codimension $k$ Faces} \label{integral}

We shall now state how, given the data of section \ref{sdata}, we can write down an
expression for the amplitude that is free from spurious poles. 
Contributions from the codimension zero faces 
$M_0^\alpha$ are straightforward to describe; we simply pull back 
\be \label{edefomega}
\omega_n = \left\langle \prod_{i=1}^K (\XX(z_i) - \p\xi(z_i) dz^i) \wedge \OO
\right\rangle_n
\ee 
to
$M_0^\alpha$ using the section $s^\alpha$ and integrate it over $M^\alpha$.
We write  $\mu_n^\alpha(m)=(s^\alpha)^*(\omega_n)$, so the contribution of $M_0^\alpha$
to the scattering amplitude is $\int_{M_0^\alpha}\mu_n^\alpha$. Since $(m;s^\alpha(m))\in X$,
this contribution is free from spurious singularities.

$\omega_n$ given in \refb{edefomega} has an important property that we shall now
describe.
Let us consider some $k$-dimensional region $P$ of $\Xi(m)$, representing the image of a
$k$-dimensional hypercube in $\RRR^K$ constructed using the 
map described in section \ref{sdata}. 
Suppose that along $P$ 
the PCO locations $z_{i_1}, \cdots z_{i_k}$ vary, keeping the other
PCO locations fixed. Suppose further that along the edge of $P$ along
which $z_{i}$ varies, its limits are $u_{i}$ and $v_{i}$. Then we have
\be \label{eintegral}
\int_P \omega_n = \pm \left\langle \prod_{s=1}^k (u_{i_s}-v_{i_s}) \prod_{j=1\atop
j\ne i_1,\cdots i_s}^K (\XX(z_i) - \p\xi(z_i) dz^i) \wedge \OO\right\rangle_{n-k}\, ,
\ee
where the overall sign has to be fixed from the orientation of the 
subspace $P$, which in turn is determined from the orientation of 
$Q_{\alpha_0\cdots\alpha_k}$
 from \refb{ebd}.
Since $(u_i,v_i,z_i)$ take values from the set
$(z_i^{(\alpha_0)}, \cdots z_i^{(\alpha_K)})$, the result is free from spurious 
singularities as
long as \refb{extracond} holds even if the subspace $P$ contains 
spurious poles. Furthermore we see that 
the result is independent of the ambiguities we have encountered in 
section \ref{sdata} in the choice of $P$,
since \refb{eintegral} has no dependence on the choices we made in constructing
$P$.

The alert reader may object to calling \refb{eintegral} an identity since the left hand side
is ill defined if $P$ contains a spurious pole of $\omega_n$. The correct viewpoint
is that we can
use \refb{eintegral} as the definition of $\int_P\omega_n$. The point however is that
we can use all the usual properties of an integral for this object, {\it e.g.} identities
of the form \refb{omox} that will be used in our analysis.
Furthermore the integral of $\omega_n$ over the two sides of \refb{ebdP} would agree.

We shall now describe how we can use this result to construct the necessary correction terms from
codimension $\ge 1$ faces of the dual triangulation.
The first non-trivial case is the contribution from codimension 1 faces.
Once a family of paths $P_{\alpha\beta}$ has been chosen for each codimension 1 face, we fiber these paths over $M_1^{\alpha\beta}$
to get a new contribution $\Sigma_{\alpha\beta}$ 
to the integration cycle that  
 ``fills the gap'' between $s^\alpha(M_0^\alpha)$ and $s^\beta(M_0^\beta)$.  To compute the contribution to the integral 
 from this new  ``vertical'' part of the integration cycle, we simply use \refb{eintegral} to
 integrate $\omega_n$ over the paths $P_{\alpha\beta}$ to reduce to an integral over $M_1^{\alpha\beta}$. By first carrying out the 
integration over $P_{\alpha\beta}$
for each $m\in M_1^{\alpha\beta}$, we can express 
$\int_{\Sigma_{\alpha\beta}}\omega_n$ as an integral over  
$M_1^{\alpha\beta}$.  The form that must be integrated over $M_1^{\alpha\beta}$ is 
\be
\mu_{n-1}^{\alpha\beta} = \int_{P_{\alpha\beta}} \omega_n.
\ee
The evaluation of the right hand side 
 can be made explicit by noting that,
along each segment of $P_{\alpha\beta}$, only one of the $z_i$'s -- say $z_j$ --
varies from an initial value $u_j$ to a final value $v_j$.
This yields the result
\be \label{yos}
\left\langle (\xi(u_j)-\xi(v_j)) \prod_{i=1\atop i\ne j}^K (\XX(z_i) - \p\xi(z_i) dz^i) \wedge \OO
\right\rangle_{n-1}\, .
\ee 
The total contribution to $\mu_{n-1}^{\alpha\beta}$ is obtained by summing over such
contributions from all the segments of $P_{\alpha\beta}$.
The results will be automatically free from spurious poles as long as \refb{extracond}
holds, since $z_i$, $u_i$, $v_i$ take values from the set $z_i^{(\alpha)}, z_i^{(\beta)}$.  We do not have
to worry about whether $P_{\alpha\beta}$ passes through the locus of spurious 
singularities 
or which path we choose from $u_i$ to $v_i$ to define it.  
The integral of the $(n-1)$-form (\ref{yos}) over $M_1^{\alpha\beta}$ has a well-defined sign,
since we have chosen orientations of each $M_1^{\alpha\beta}$.

Let us now consider a codimension two face 
$M_2^{\alpha\beta\gamma}$  that is shared by $M_0^\alpha$, $M_0^\beta$ and
$M_0^\gamma$. On $M_1^{\alpha\beta}$, the integral in the vertical direction is carried
out over a particular path $P_{\alpha\beta}$ 
connecting $s^\alpha(m)$ to $s^\beta(m)$, and the analog is true for 
$M_1^{\beta\gamma}$ and $M_1^{\gamma\alpha}$. Now if it so happens that on 
$M_2^{\alpha\beta\gamma}$, 
$P_{\alpha\beta}$, $P_{\beta\gamma}$ and $P_{\gamma\alpha}$ together 
describe zero path (i.e.\ $P_{\alpha\beta}+P_{\beta\gamma}=-P_{\gamma\alpha}$)
then we do not need any correction term on $M_2^{\alpha\beta\gamma}$, since the integration cycle has no gap. However
generically $P_{\alpha\beta}$, $P_{\beta\gamma}$ and $P_{\gamma\alpha}$ will describe
a closed path in $\Xi$, leaving a gap in the integration cycle in $Y$, and we need 
to fill the gap
by including, for each $m\in M_2^{\alpha\beta\gamma}$, a two-dimensional vertical
segment that represents a two-dimensional subspace 
of $\Xi$ bounded by $P_{\alpha\beta}$,
$P_{\beta\gamma}$ and $P_{\gamma\alpha}$. 
We choose this to be the subspace $P_{\alpha\beta\gamma}$ constructed in section
\ref{sdata}. We now add to the integration cycle the spaces 
$\Sigma_{\alpha\beta\gamma}$ obtained by  
fibering $P_{\alpha\beta\gamma}$ over $M_2^{\alpha\beta\gamma}$,  and
explicitly carry out integration
over $P_{\alpha\beta\gamma}$ for a given point $m\in M_2^{\alpha\beta\gamma}$ to
get a form
\be\label{buffalo}
\mu_{n-2}^{\alpha\beta\gamma} \equiv \int_{P_{\alpha\beta\gamma}}\omega_n\, ,
\ee
which then has to be integrated over $M_2^{\alpha\beta\gamma}$ to get the cxdimension 2 correction. Again since 
$P_{\alpha\beta\gamma}$ is expressed as a sum of rectangles and along each rectangle
only two of the PCO locations vary,
contribution from each rectangle can be computed   
using the general form described in \refb{eintegral}.
The condition \refb{extracond} ensures that $\mu_{n-2}^{\alpha\beta\gamma}$ is  
well-defined, not affected
by spurious poles. 

This continues to higher order. For example, at a codimension 3 face $M_3^{\alpha\beta
\gamma\delta}$, four codimension 2 faces $M_{\alpha\beta\gamma}$, 
$M_{\beta\gamma\delta}$, $M_{\gamma\delta\alpha}$ and $M_{\delta\alpha\beta}$
meet. Associated with them are two-dimensional subspaces $P_{\alpha\beta\gamma}$,
$P_{\beta\gamma\delta}$, $P_{\gamma\delta\alpha}$ and $P_{\delta\alpha\beta}$
of $\Xi$.
Together they describe a two-dimensional closed subspace of $\Xi$ and hence 
can be taken to be the boundary of a three-dimensional subspace of $\Xi$.\footnote{For
reasons explained earlier, there is no topological obstruction. We can always work
with the subspaces  $Q_{\alpha\beta\gamma}$,
$Q_{\beta\gamma\delta}$, $Q_{\gamma\delta\alpha}$ and $Q_{\delta\alpha\beta}$
of $\RRR^K$, find the subspace $Q_{\alpha\beta\gamma\delta}$ bounded by them, and then
map it back to $\Xi$ to get $P_{\alpha\beta\gamma\delta}$.}
We take this to be the subspace $P_{\alpha\beta\gamma\delta}$ introduced in section \ref{sdata}
and fill the gap in the integration cycle by adding to it the space 
$\Sigma_{\alpha\beta\gamma\delta}$ obtained by fibering 
$P_{\alpha\beta\gamma\delta}$
over $M_3^{\alpha\beta\gamma\delta}$. The integration over $P_{\alpha\beta\gamma\delta}$
can be performed explicitly for each 
$m\in M_3^{\alpha\beta\gamma\delta}$ using \refb{eintegral},  
yielding a result $\mu_{n-3}^{\alpha\beta\gamma\delta}$ free from
spurious poles, which can then be integrated over $M_3^{\alpha\beta\gamma\delta}$.

At the end, the full amplitude may be expressed as
\be \label{efull}
\sum_{k=0}^K (-1)^{k(k+1)/2}
\sum_{\{\alpha_0,\cdots, \alpha_k\}} \int_{M_k^{\alpha_0\cdots \alpha_k}}
\mu_{n-k}^{\alpha_0\cdots \alpha_k}\, ,
\ee
with 
\be \label{edefmunk}  
\mu_{n-k}^{\alpha_0\cdots\alpha_k} = \int_{P_{\alpha_0\cdots\alpha_k}}\omega_n
\ee
 being an $n-k$ form on 
$M_k^{\alpha_0\cdots \alpha_k}$ that is free from spurious poles.

If the vertical paths could be chosen consistently to make the $\simeq$ symbol in \refb{ebdP}  an equality, and if there
were no spurious poles, then \refb{efull} could be interpreted as the integral of
$\omega_n$ over a continuous integration cycle obtained by joining the sections
$\{s^\alpha\}$ over $\{\UU_\alpha\}$ and the vertical segments 
$\{P_{\alpha_0\cdots\alpha_k}\}$. 
Even though all this is not true, the expression \refb{efull} shares 
all the necessary properties of an amplitude that would be obtained by integrating
$\omega_n$ over a continuous integration cycle. In particular the amplitude is
gauge invariant -- if any of the external states is a BRST trivial state then $\omega_n$
is an exact form and its integral vanishes (as long as the contribution from the boundary
of the moduli space vanishes).  Similarly if we had made a different choice of the sections
$\{s^\alpha\}$ or a different choice of the paths $Q_{\alpha_0\cdots \alpha_n}$ then it
would correspond to a different choice of the integration cycle that is homologous to the
original cycle, but the result for the
amplitude remains unchanged since $\omega_n$ is closed.  
We shall give more explicit proofs of these properties in sections 
\ref{sequiv}-\ref{decoupling}.   These proofs will make clear the need for the peculiar minus sign in eqn. (\ref{efull}).

\subsection{An Example}

We shall now illustrate the above method by explicitly
constructing the integrands on codimension 1 and 2 faces 
for three PCO's. To simplify notation, let us define
\be\label{edefyy}
\YY(z) = \XX(z) - \p\xi(z) dz\, .
\ee
Thus the form that must  be integrated over the codimension
zero face $M_0^\alpha$ is
\be 
\mu_n^\alpha(m)= \left\langle \YY(z_1^{(\alpha)}) \YY(z_2^{(\alpha)}) \YY(z_3^{(\alpha)}) 
\, \OO\right \rangle_n\, ,
\ee
where all products are to be interpreted as wedge products.
In the following we shall work on three codimension zero faces labelled $a$,
$b$ and $c$ and determine $\mu_{n-1}^{\alpha\beta}$ and 
$\mu_{n-2}^{\alpha\beta\gamma}$ for $\alpha,\beta,\gamma$ taking values $a$,
$b$, $c$.

We begin with the construction of $\mu_{n-1}^{ab}(m)$. 
We represent the sections $s^a=
(z^{(a)}_1,z^{(a)}_2,z^{(a)}_3)$ and $s^b=(z^{(b)}_1,z^{(b)}_2,z^{(b)}_3)$ 
as the opposite  corners (0,0,0) and (1,1,1) of a unit cube. 
The interpretation of the other corner points has been described earlier.
Now we have to ``fill the
gap'' between the two opposite corners by choosing a path $Q_{ab}$ between them along the edges of the cube. 
Let us take this to consist of straight line segments traversing the path
(0,0,0)-(1,0,0)-(1,1,0)-(1,1,1), corresponding to moving
first $z_1$, then $z_2$, and finally $z_3$. $\mu_{n-1}$ is then given by the
integral of $\YY(z_1)\YY(z_2) \YY(z_3)$
along this curve.
Note that  the integral could run into spurious 
poles along the way; so at this stage we still regard this as a formal expression
or
a bookkeeping device for generating the $\mu_{n-1}^{ab}$.
In any case, using \refb{edefyy} we get the result of integral to be
\ben \label{e109}
\mu_{n-1}^{ab}&=& \bigg\langle \bigg[
(\xi(z^{(a)}_1)-\xi(z^{(b)}_1)) \YY(z^{(a)}_2) \YY(z^{(a)}_3) + (\xi(z^{(a)}_2)-\xi(z^{(b)}_2)) 
\YY(z^{(b)}_1) \YY(z^{(a)}_3) \nonumber \\
&& \qquad 
+ (\xi(z^{(a)}_3)-\xi(z^{(b)}_3)) \YY(z^{(b)}_1) \YY(z^{(b)}_2)\bigg] \wedge \OO\bigg
\rangle_{n-1}\, .
\een
This depends only on the corner points and is free from any singularity. If we had
chosen a different path connecting (0,0,0) and (1,1,1) we would get a different
result that is cohomologically equivalent to the one given above. 
$\mu_{n-1}^{ba}(m)$
will be the negative of \refb{e109} and {\it not what is obtained by exchanging $a$
and $b$ in eqn. \refb{e109}} since the result depends on the choice of a  path between the two opposite corners.   To compute
$\mu_{n-1}^{ba}(m)$, we have to reverse the order in which the PCO's are moved, leading to $\mu_{n-1}^{ba}(m)=-\mu_{n-1}^{ab}(m)$.

We also define $\mu_{n-1}^{bc}(m)$ and
$\mu_{n-1}^{ca}(m)$ similarly, i.e. by choosing paths 
$Q_{bc}$ and $Q_{ca}$ and
integrating $\omega_n$
along the images of these paths in $\Xi$. 
For definiteness, we shall assume that these paths follow the
same ordering conventions as the ones used to defining $\mu_{n-1}^{ab}$, i.e.\
we move $z_1$ first, then $z_2$ and then $z_3$.
This is not necessary -- one could have chosen any other ordering
prescription for these paths independently of how we have chosen $Q_{ab}$. 
With this choice we get
\ben \label{e109a}
\mu_{n-1}^{bc}&=& \bigg\langle \bigg[
(\xi(z^{(b)}_1)-\xi(z^{(c)}_1)) \YY(z^{(b)}_2) \YY(z^{(b)}_3) + (\xi(z^{(b)}_2)-\xi(z^{(c)}_2)) 
\YY(z^{(c)}_1) \YY(z^{(b)}_3) \nonumber \\
&& \qquad 
+ (\xi(z^{(b)}_3)-\xi(z^{(c)}_3)) \YY(z^{(c)}_1) \YY(z^{(c)}_2)\bigg] \wedge \OO\bigg
\rangle_{n-1}\, , \nonumber \\
\mu_{n-1}^{ca}&=& \bigg\langle \bigg[
(\xi(z^{(c)}_1)-\xi(z^{(a)}_1)) \YY(z^{(c)}_2) \YY(z^{(c)}_3) + (\xi(z^{(c)}_2)-\xi(z^{(a)}_2)) 
\YY(z^{(a)}_1) \YY(z^{(c)}_3) \nonumber \\
&& \qquad 
+ (\xi(z^{(c)}_3)-\xi(z^{(a)}_3)) \YY(z^{(a)}_1) \YY(z^{(a)}_2)\bigg] \wedge \OO\bigg
\rangle_{n-1}\, .
\een

Now let us turn to $\mu_{n-2}^{abc}(m)$.
In the spirit of the algorithm described earlier, we represent 
$(z^{(a)}_1,z^{(a)}_2,z^{(a)}_3)$, $(z^{(b)}_1,z^{(b)}_2,z^{(b)}_3)$ and
$(z^{(c)}_1,z^{(c)}_2,z^{(c)}_3)$
as the points $(0,0,0)$, $(1,1,1)$ and $(2,2,2)$ in $\RRR^3$ respectively, and interpret
other integer points accordingly.
In this representation $Q_{ab}$ describes a path connecting (0,0,0) to (1,1,1),
$Q_{bc}$ describes a path connecting
(1,1,1) to (2,2,2) and  $Q_{ca}$ describes a path connecting
(2,2,2) to (0,0,0).
Together the three paths $Q_{ab}$, $Q_{bc}$ and $Q_{ca}$
describe a closed curve in $\RRR^3$
traversing  the links of the lattice of integers.  By gluing these three paths together end to end,
we make a closed path from the origin in $\RRR^3$ to itself.  With the choice just described, this closed
path is
\ben \label{eline}
&& (0,0,0) - (1,0,0) - (1,1,0) - (1,1,1) - (2,1,1) - (2,2,1) \nonumber \\
&& - (2,2,2) - (0,2,2) - (0,0,2) - (0,0,0)\, .
\een
Now the general algorithm instructs us to find a surface $-Q_{abc}$
enclosed by $Q_{ab}+Q_{bc}+Q_{ca}$,
and consisting of a union of unit squares whose corners are lattice points. 
Equivalently, we can use rectangles built by gluing together such unit squares.
We then
find the image $P_{abc}$ of $Q_{abc}$ in $\Xi$ and integrate
$\YY(z_1)\YY(z_2) \YY(z_3)$
over this two-dimensional space to find $\mu_{n-2}^{abc}$. 
Again the surface runs
through spurious poles but 
we can regard this as a formal integral or a bookkeeping device for
generating $\mu_{n-2}^{abc}$, 
which will eventually be expressed in terms of only the
corner points of the rectangles. 
There are many ways of choosing $Q_{abc}$; we will describe a specific choice.
We shall specify each rectangle by giving its corner points,
and then give the result of the integral over the image of the rectangle in $\Xi$:
\ben \label{e111}
(0,1,2) - (2,1,2) - (2,2,2) - (0,2,2) &:&  (\xi(z_1^{(c)}) 
- \xi(z_1^{(a)})) (\xi(z_2^{(c)})-\xi(z_2^{(b)}))\YY(z_3^{(c)}) 
\nonumber \\
(0,0,2) - (1,0,2) - (1,1,2) - (0,1,2) &:&(\xi(z_1^{(b)}) 
- \xi(z_1^{(a)})) (\xi(z_2^{(b)}) - \xi(z_2^{(a)})) \YY(z_3^{(c)}) 
\nonumber \\
(0,0,0) - (1,0,0) - (1,0,2) - (0,0,2) &:&(\xi(z_1^{(b)}) 
- \xi(z_1^{(a)})) (\xi(z_3^{(c)}) - \xi(z_3^{(a)})) \YY(z_2^{(a)}) 
\nonumber \\
(1,0,0) - (1,1,0) - (1,1,2) - (1,0,2) &:&(\xi(z_2^{(b)}) 
- \xi(z_2^{(a)})) (\xi(z_3^{(c)}) - \xi(z_3^{(a)})) \YY(z_1^{(b)}) 
\nonumber \\
(1,1,1) - (2,1,1) - (2,1,2) - (1,1,2)  &:&(\xi(z_1^{(c)}) 
- \xi(z_1^{(b)})) (\xi(z_3^{(c)}) - \xi(z_3^{(b)})) \YY(z_2^{(b)}) 
\nonumber \\
(2,1,1) - (2,2,1) - (2,2,2) - (2,1,2) &:&(\xi(z_2^{(c)}) 
- \xi(z_2^{(b)})) (\xi(z_3^{(c)}) - \xi(z_3^{(b)})) \YY(z_1^{(c)}) 
\nonumber \\
\een
If we denote the sum of all these terms by $\AAA$ then
$\mu_{n-2}^{abc}(m)$ 
will be given by 
\be
\mu_{n-2}^{abc}(m) = -\left\langle \AAA\wedge \OO\right\rangle_{n-2}
\ee
where the extra minus sign reflects the fact that 
$Q_{abc}$ is enclosed by  
$-(Q_{ab}+Q_{bc}+Q_{ca})$. 

Of course, one can construct many other candidates for
$\mu_{n-2}^{abc}$ by choosing a different set
of rectangles which have the path 
\refb{eline} as boundary, but we shall argue in
section \ref{sequiv} that they will all give the same result for the final integral.

\subsection{Avoiding the Spurious Singularities by Fine Coverings} \label{ssection}

We now turn to the proof that, for a sufficiently fine dual triangulation $\Upsilon$, it is possible to pick local sections that satisfy the condition
\refb{extracond}. We shall in fact prove a slightly more general
statement.  Given any positive integer $t$, and given any covering of  $M$ by sufficiently small open sets $\UU_\alpha$
any one of which has non-empty intersection with at most $t$ others, 
we can choose sections $s^\alpha:\UU_\alpha\to X$ with the property that
\be \label{ereqdrep}
(m;s^{\alpha_0}(m), \cdots ,s^{\alpha_k}(m))\in X^{(k+1)} \quad 
\hbox{for $m\in \UU_{\alpha_0}\cap \cdots \cap \UU_{\alpha_k}$}\, .
\ee
  This implies the condition (\ref{extracond}) if the dual triangulation $\Upsilon$ is fine enough. (We choose the $\UU_\alpha$
to be open sets each of which is a slight thickening of one of the codimension zero polyhedra in $\Upsilon$.)  

We start with a preliminary about Riemann surfaces.   If $\UU\subset M$ is a sufficiently small open set, we can think of the surfaces $\Sigma(m)$, $m\in \UU$
as a constant family of two-manifolds (oriented and with punctures) with only the complex structure depending on $m$.  There is no natural way to do
this, and we simply pick any way.

Once this is done for each $\UU_\alpha$, we can pick the sections $s_\alpha:\UU_\alpha\to X$ to be ``constant.'' What this means is that we pick a base point $m_\alpha$
in each $\UU_\alpha$ and we choose $s^\alpha$ (a collection of punctures in $\Sigma(m_\alpha)$ at which PCO's are to be inserted)
at the point $m_\alpha$.  Then for $m_\alpha'\in \UU_\alpha$, since we have picked an identification of
$\Sigma(m_\alpha')$ with $\Sigma(m_\alpha)$, we just define $s^\alpha$ by choosing the ``same''  PCO insertion points on $\Sigma(m_\alpha')$ as on $\Sigma(m_\alpha)$.  

Will a section defined this way avoid the locus of spurious singularities in $\Xi(m_\alpha)=\Sigma(m_\alpha)\times \dots\times \Sigma(m_\alpha)$?  Let us call the bad locus 
$\Xi_0(m_\alpha)$; it is of real
codimension 2 in $\Xi(m_\alpha)$.  A spurious singularity is avoided at $m_\alpha$ if $s^\alpha\notin \Xi_0(m_\alpha)$.  To avoid a spurious singularity from occurring at any $m_\alpha'\in\UU_\alpha$,
we want $s^\alpha$ to be sufficiently far from $\Xi_0(m_\alpha)$.   For $\veps$ small and positive, using
some arbitrary metric on $\Sigma(m_\alpha)$, let $\Xi_0^\veps(m_\alpha)$ be a tube centered at $\Xi_0$ of radius $\veps$.  Then if $\UU_\alpha$ 
is small enough, the ``constant'' section $s^\alpha$
avoids spurious singularities for all $m_\alpha'\in\UU_\alpha$ provided that $s^\alpha\notin \Xi_0^\veps(m_\alpha)$.  

We observe that the volume of $\Xi_0^\veps(m_\alpha)$ is of order $\veps^n$ (where $n=\mathrm{dim}\,M$) and in particular for given positive integer $T$ and small enough $\veps$, the union
of $T$ tubes such as $\Xi_0^\veps$ covers only a small part of $\Xi$.

Now suppose we are given a covering of $M$ by sufficiently small open sets $\UU_\alpha$, each of which intersects at most $t$ others.  One at a time, we pick one of
the $\UU_\alpha$ and select a base point $m_\alpha$ and a section $s^\alpha:\UU_\alpha\to X$ that is constant in the above sense.  To satisfy \refb{ereqdrep}
(where we consider only those $\UU_\beta$ for which $s^\beta$ has already been chosen),  $s^\alpha$ must be chosen to avoid at most $T$ tubes similar to $\Xi_0^\veps(m_\alpha)$,
where $T$ is a positive integer that depends only on $t$.  ($T$ exceeds $t$ because there are many conditions to satisfy in \refb{ereqdrep}.)  For small enough open sets
$\UU_\alpha$ and therefore small enough $\veps$, there is no obstruction to doing this at any stage.

\section{Dependence on the Choice of the Vertical Segments} \label{sequiv}

We have seen that the definition of the string amplitude using the prescription for
vertical integration suffers from ambiguities since the choice of the subspaces 
$Q_{\alpha_0\cdots \alpha_k}$ of $\RRR^K$ have some freedom and as a 
consequence their images $P_{\alpha_0\cdots\alpha_k}$
in $\Xi$ also enjoy the same freedom. Thus we need to
show that the result for the amplitude is independent of this choice. This is what we
shall show in this section.

If there were no spurious poles, and if there were no topological obstruction to the choice
of a smooth section $s:M\to Y$ of the projection $\vp:Y\to M$, then a scattering amplitude
would be defined simply as $\int_M s^*(\omega_n)$.  Then Stoke's theorem together with the 
fact that $d \omega_n=0$ would imply that the scattering amplitude is independent of the choice of $s$
within its homology class.  One would still have to worry about a possible dependence on the homology class of $s$.

Actually, there are spurious poles and a global section $s:M\to Y$ very likely does not exist. 
We have avoided both issues in this paper by using a piecewise construction based on local sections $s^\alpha:\UU_\alpha\to X\subset Y$
that avoid spurious singularities.  Formally the integration cycle then contains ``vertical'' segments, but since one does not
have to really pick specific vertical segments, spurious singularities do not enter, there is no need to construct a global section of $X\to M$ or
even of $Y\to M$, and there is no issue concerning the homology class of such a section.

However, we do want to show that the amplitude that we have defined is independent of the choices that were made.
We focus on the situation where  two choices of integration cycle correspond to the
same dual triangulation and the same 
$\{s^\alpha\}$ but different $\{Q_{\alpha_0\cdots \alpha_k}\}$,
-- the case where the choice of dual triangulation or the choice of 
$\{s^{\alpha}\}$'s change will be discussed in a
somewhat more general context in section \ref{smooth}. 
We denote
by $Q_{\alpha_0\cdots \alpha_k}$ and $\t Q_{\alpha_0\cdots \alpha_k}$ the two sets
of choices for these subspaces. Now since (for example)
$Q_{\alpha\beta}$ and $\t Q_{\alpha\beta}$ are paths in $\RRR^K$ with
the same end points, we have
\be 
\p Q_{\alpha\beta} - \p \t Q_{\alpha\beta} = 0\, .
\ee
This allows us to write
\be \label{es1}
Q_{\alpha\beta} - \t Q_{\alpha\beta} = \p V_{\alpha\beta}\, ,
\ee
where $V_{\alpha\beta}$ is a two-dimensional subspace of $\RRR^K$. We take this
to be composed of rectangles lying along coordinate planes just as we did for
$Q_{\alpha\beta\gamma}$.  We make the choices so that $V_{\beta\alpha}=-V_{\alpha\beta}$ (where $-V_{\alpha\beta}$ is
$V_{\alpha\beta}$ with opposite orientation).

Next we note, using \refb{ebd}, its counterpart involving $\t Q$'s, and \refb{es1} that
\be \label{es2}
\p Q_{\alpha\beta\gamma} - \p \t Q_{\alpha\beta\gamma} 
= - (\p V_{\alpha\beta} + \p V_{\beta\gamma} + \p V_{\gamma\alpha} )
\ee
This allows us to construct a three dimensional subspace $V_{\alpha\beta\gamma}$ of
$\RRR^K$ satisfying
\be \label{e28}
Q_{\alpha\beta\gamma} - \t Q_{\alpha\beta\gamma}
= - (V_{\alpha\beta} + V_{\beta\gamma} + V_{\gamma\alpha} )
+ \p V_{\alpha\beta\gamma}
\ee
Again we shall choose $V_{\alpha\beta\gamma}$ to be composed of hypercuboids
 whose
three sides lie along the three coordinate axes, and antisymmetric under the exchange
of $\alpha,\beta,\gamma$. 

At the next step we use
\ben \label{e29}
\p Q_{\alpha\beta\gamma\delta} - \p\t Q_{\alpha\beta\gamma\delta}
&=& - \{ (Q_{\alpha\beta\gamma} - \t Q_{\alpha\beta\gamma}) 
- (Q_{\alpha\beta\delta} - \t Q_{\alpha\beta\delta})  \nonumber \\
&& +
(Q_{\alpha\gamma\delta} - \t Q_{\alpha\gamma\delta}) -
(Q_{\beta\gamma\delta} - \t Q_{\beta\gamma\delta}) \} \nonumber \\
&=& - \p (V_{\alpha\beta\gamma} - V_{\alpha\beta\delta} +V_{\alpha\gamma\delta}
- V_{\beta\gamma\delta})\, .
\een
The terms involving
$V_{\alpha\beta}$, etc., have canceled in arriving at this result. We can now find a $V_{\alpha\beta\gamma\delta}$
such that
\be 
Q_{\alpha\beta\gamma\delta} - \t Q_{\alpha\beta\gamma\delta}
= - (V_{\alpha\beta\gamma} - V_{\alpha\beta\delta} +V_{\alpha\gamma\delta}
- V_{\beta\gamma\delta}) +\p V_{\alpha\beta\gamma\delta}\, .
\ee

The generalization is now obvious. We get 
\be 
\p (Q_{\alpha_0\cdots \alpha_k} - \t Q_{\alpha_0\cdots \alpha_k})
= - \sum_{i=0}^k (-1)^{k-i} \p V_{\alpha_0\cdots \alpha_{i-1}\alpha_{i+1}
\cdots \alpha_k}\, ,
\ee
and hence we can find $V_{\alpha_0\cdots \alpha_k}$
satisfying
\be \label{eimage}
Q_{\alpha_0\cdots \alpha_k}  - \t Q_{\alpha_0\cdots \alpha_k} = - 
\sum_{i=0}^k (-1)^{k-i} V_{\alpha_0\cdots \alpha_{i-1}\alpha_{i+1}
\cdots \alpha_k} + \p V_{\alpha_0\cdots \alpha_k}
\ee
Furthermore we can choose $V_{\alpha_0\cdots \alpha_k}$ to be totally antisymmetric in the 
labels $\alpha_0,\cdots \alpha_k$ and to be composed of $(k+1)$-dimensional
hypercuboids whose sides lie along coordinate axes.

In section \ref{amplitude}, we  
picked maps from $Q_{\alpha_0\cdots
\alpha_k}$ to $\Xi(m)$  for $m\in M_k^{\alpha_0\cdots\alpha_k}$ 
and integrated $\omega_n$ over the image, which we called
$P_{\alpha_0\cdots \alpha_k}$.  This construction was formal since the 
$P_{\alpha_0\cdots \alpha_k}$'s could pass through spurious poles and also there could
be many topologically different choices for $P_{\alpha_0\cdots\alpha_k}$.
But the definitions were made so that the
integral over $P_{\alpha_0\cdots\alpha_k}$ could be defined as in eqn. 
(\ref{eintegral}) (for
example) without really having to pick the map from $Q_{\alpha_0\cdots \alpha_k}$ to $\Xi(m)$.

 In a similar spirit,
we formally extend the maps from $Q_{\alpha_0\cdots \alpha_k}$ and $\t Q_{\alpha_0\cdots\alpha_k}$
to $\Xi(m)$ to maps
from  $V_{\alpha_0\cdots \alpha_k}$ to $\Xi(m)$.  
Formally, we let $U_{\alpha_0\cdots \alpha_k}$ be the image of $V_{\alpha_0\cdots
\alpha_k}$ in $\Xi$ and define an $(n-k-1)$-form
$\chi_{n-k-1}^{\alpha_0\cdots \alpha_k}$ on $M_k^{\alpha_0\cdots \alpha_k}$ by integration
over $U_{\alpha_0\cdots\alpha_k}$:
\be \label{edefchi}
\chi_{n-k-1}^{\alpha_0\cdots \alpha_k} = \int_{U_{\alpha_0\cdots \alpha_k}} \omega_n\, .
\ee
Just as in section \ref{amplitude}, this is a symbolic formula.  We really define $\chi_{n-k-1}^{\alpha_0
\cdots\alpha_k}$ by a conformal field theory formula analogous to 
eqn. (\ref{eintegral}).
 
Now the image of \refb{eimage} in $\Xi$ gives
\be
P_{\alpha_0\cdots \alpha_k} - \t P_{\alpha_0\cdots \alpha_k}
\simeq - 
\sum_{i=0}^k (-1)^{k-i} U_{\alpha_0\cdots \alpha_{i-1}\alpha_{i+1}
\cdots \alpha_k} + \p U_{\alpha_0\cdots \alpha_k}\, ,
\ee
where $\simeq$ has the same interpretation as in \refb{ebdP}.
Hence on $M_{\alpha_0\cdots \alpha_k}$,
\ben
\mu_{n-k}^{\alpha_0\cdots \alpha_k} -\t\mu_{n-k}^{\alpha_0\cdots \alpha_k} 
&=& \int_{P_{\alpha_0\cdots \alpha_k}} \omega_n
- \int_{\t P_{\alpha_0\cdots \alpha_k}} \omega_n \nonumber \\
&=&- 
\sum_{i=0}^k (-1)^{k-i} \int_{U_{\alpha_0\cdots \alpha_{i-1}\alpha_{i+1}
\cdots \alpha_k}} \omega_n+ \int_{\p U_{\alpha_0\cdots \alpha_k}} \omega_n
\, .
\een
Now for any $p$-form 
$\Omega_p$ on $Y$, 
and an $\ell$-dimensional subspace
$R_\ell(m)$ of $\Xi(m)$ defined on a local neighbourhood $\UU$ of $M$,
we have 
\be\label{omox}\int_{\partial R_\ell}
\Omega_p=\int_{R_\ell}d\, \Omega_p - (-1)^\ell  
d\int_{R_\ell}\Omega_p. \ee
This is a relation among $(p-\ell+1)$-forms on $\UU$.  
 In our analysis we shall apply this identity for $\Omega_p=\omega_n$ 
 which has spurious singularities. Nevertheless the identity holds with the
 definition of the various integrals as given in \refb{eintegral}.
For $p=n$, $\Omega_p=\omega_n$, $\ell=k+1$  
and $R_\ell=U_{\alpha_0\cdots\alpha_k}$,
one term drops out since $d\omega_n=0$.
Hence
\be
\mu_{n-k}^{\alpha_0\cdots \alpha_k} -\t\mu_{n-k}^{\alpha_0\cdots \alpha_k}  =- 
\sum_{i=0}^k (-1)^{k-i} \int_{U_{\alpha_0\cdots \alpha_{i-1}\alpha_{i+1}
\cdots \alpha_k}} \omega_n-  (-1)^{k+1}  
d\int_{U_{\alpha_0\cdots \alpha_k}} \omega_n\, .
\ee
Using the definition \refb{edefchi} we get, on $M_k^{\alpha_0\cdots \alpha_k}$
\be \label{edefre}
\mu_{n-k}^{\alpha_0\cdots \alpha_k} -\t\mu_{n-k}^{\alpha_0\cdots \alpha_k}  
=- 
\sum_{i=0}^k (-1)^{k-i} \chi_{n-k}^{\alpha_0\cdots \alpha_{i-1}\alpha_{i+1}
\cdots \alpha_k} - (-1)^{k+1}   
d\chi_{n-k-1}^{\alpha_0\cdots \alpha_k} \, .
\ee

Let us now examine the difference between the full amplitudes computed using
the $Q_{\alpha_0\cdots\alpha_k}$'s and the $\t Q_{\alpha_0\cdots\alpha_k}$'s.
Using \refb{efull} this is given by  
\ben \label{efin}
&& \sum_{k=0}^{K} (-1)^{k(k+1)/2} \sum_{\{\alpha_0\cdots \alpha_k\}}
\int_{M_{k}^{\alpha_0\cdots \alpha_k}} (\mu_{n-k}^{\alpha_0\cdots \alpha_k} 
-\t\mu_{n-k}^{\alpha_0\cdots \alpha_k}) \nonumber \\
&=&  -\sum_{k=0}^{K} (-1)^{k(k+1)/2}  
\sum_{\{\alpha_0\cdots \alpha_k\}} \sum_{i=0}^k (-1)^{k-i} 
\int_{M_{k}^{\alpha_0\cdots  \alpha_k} }
\chi_{n-k}^{\alpha_0\cdots \alpha_{i-1}\alpha_{i+1}
\cdots \alpha_k} \nonumber \\
&& +\sum_{k=0}^{K} (-1)^{k(k+1)/2 + (k+1)} \,
\sum_{\{\alpha_0\cdots \alpha_k\}} 
\sum_\beta \int_{M_{k+1}^{\alpha_0\cdots \alpha_k\beta} }
\chi_{n-k-1}^{\alpha_0\cdots \alpha_k} \nonumber \\
\een
where we have used \refb{edefre} and manipulated the second term using
\be \label{efin1}
\int_{M_{k}^{\alpha_0\cdots \alpha_k}}  d 
\chi_{n-k-1}^{\alpha_0\cdots \alpha_k}
= \int_{\p M_{k}^{\alpha_0\cdots \alpha_k}}  \chi_{n-k-1}^{\alpha_0\cdots \alpha_k}
= -\sum_\beta \int_{M_{k}^{\alpha_0\cdots \alpha_k\beta}}  
\chi_{n-k-1}^{\alpha_0\cdots \alpha_k}\, ,
\ee
using \refb{eorientation}. The sum over $\beta$ in \refb{efin}, \refb{efin1} 
run over all $\beta\neq \alpha_0,\cdots \alpha_k$  
for which
$M_0^\beta$ overlaps with 
$M_{k}^{\alpha_0\cdots \alpha_k}$.
It is now easy to see that the terms in \refb{efin} cancel pairwise, making the result
vanish. For this it is important that we have the $(-1)^{k(k+1)/2}$ factor in the
summand in \refb{efull}.  

\section{Smooth Measure} \label{smooth} 

The integration measure on $M$ that we have constructed in section \ref{triangulations} is
not smooth since it is discontinuous across the boundaries separating a pair of 
codimension zero faces and we have to add correction terms on codimension 1
faces to compensate for this discontinuity. We shall now describe an alternate procedure that
constructs a smooth integration measure on $M$.

This requires the following ingredients.
\begin{enumerate}
\item Choose a sufficiently fine cover of $M$ by open sets 
$\{\UU_\alpha\}$ so that on each open set 
we can choose a local section $s^\alpha$ of $X$ satisfying \refb{ereqdrep}:
\be \label{ereqdreprep}
(m;s^{\alpha_0}(m), \cdots s^{\alpha_k}(m))\in X^{(k+1)} \quad 
\hbox{for $m\in \UU_{\alpha_0}\cap \cdots \cap \UU_{\alpha_k}$}\, .
\ee
\item On each overlap $\UU_\alpha\cap\UU_\beta$ we choose a path
$Q_{\alpha\beta}$ in $\RRR^K$ (and hence its image $P_{\alpha\beta}$ in
$\Xi$) as in section \ref{sdata}. Similarly, on each triple overlap $\UU_\alpha\cap
\UU_\beta\cap \UU_\gamma$, we choose a surface $Q_{\alpha\beta\gamma}$ in
$\RRR^K$ (and its image $P_{\alpha\beta\gamma}$ in $\Xi$) satisfying \refb{ebd}.
We continue this and choose $Q_{\alpha_0\cdots \alpha_k}\subset \RRR^K$
for all $k$ up to $K$, satisfying \refb{ebdfull}:
\be \label{ebdfullrep}
\p Q_{\alpha_0\cdots \alpha_k} = -\sum_{i=0}^k (-1)^{k-i} 
Q_{\alpha_0, \cdots \alpha_{i-1}, \alpha_{i+1}, \cdots \alpha_k}
\, .
\ee
\item Using the image $P_{\alpha_0\cdots \alpha_k}$ of $Q_{\alpha_0\cdots \alpha_k}$
in $\Xi$, we now follow the procedure of section \ref{integral} to construct the
$n-k$ form $\mu_{n-k}^{\alpha_0\cdots \alpha_k}$ for all sets 
$\{\alpha_0,\cdots \alpha_k\}$ and all $k$ from 0 to $K$ for which 
$\UU_{\alpha_0}\cap\cdots \cap \UU_{\alpha_k}$ is non-empty.
The difference with the case analyzed in section \ref{integral} is that
$\mu_{n-k}^{\alpha_0\cdots \alpha_k}$ is now defined on the open set
$\UU_{\alpha_0}\cap\cdots \cap\UU_{\alpha_k}$ instead of just on a 
codimension $k$ subspace.
\item We now choose a partition of unity subordinate to the open cover $\UU_\alpha$.  This means that we   choose, on each $\UU_\alpha$, a 
smooth function $A^{(\alpha)}(m)$ satisfying
\be \label{econsA}
A^{(\alpha)}(m) = 0 \quad \hbox{for $m\not\in \UU_\alpha$}, \quad 
\sum_\alpha A^{(\alpha)}(m) = 1\, .
\ee
\end{enumerate}
We are now ready to write down the expression for the amplitude generalizing
\refb{efull}:
\be \label{esmooth}
\AAA=\int_M \sum_{k=0}^K (-1)^{k(k-1)/2} \sum_{\{\alpha_0,\cdots, \alpha_k\}} A^{(\alpha_0)} 
d A^{(\alpha_1)}\wedge \cdots \wedge dA^{(\alpha_k)} \wedge 
\mu_{n-k}^{\alpha_0\cdots\alpha_k}\, .
\ee
The sum over each $\alpha_i$ runs over all the open sets in the cover, but due to the presence
of the $A^{\alpha_i}$'s in the summand and the antisymmetry of $\mu$, 
we only pick up a non-zero contribution
from those combinations $\{\alpha_0,\cdots\alpha_k\}$ 
for which $\alpha_i$'s are all different and
the sets $\{\UU_{\alpha_i}\}$ for $0\le i\le k$ have
an overlap.
In order to prove that \refb{esmooth} 
is a sensible expression for the amplitude, we have to show that
\begin{enumerate}
\item It is independent of the choice of $Q_{\alpha_0\cdots \alpha_k}$.
\item It is independent of the choice of the $A^{(\alpha)}$'s.
\item It is independent of the choice of the sections $\{s^\alpha\}$.  
\item It is independent of the choice of the open cover.
\item In an appropriate limit, it reduces to \refb{efull}.
\end{enumerate}
It is also necessary to show gauge invariance, but we postpone this to section \ref{decoupling}.

We begin with the proof of the first property. If $\mu$ and $\tilde\mu$ denote the
$\mu$'s associated with two different choices of 
$Q_{\alpha_0\cdots\alpha_k}$, then their  
difference can be expressed as in \refb{edefre}. Using this we get the difference between
the two amplitudes to be
\ben
\Delta &=& - \int_M \sum_{k=0}^K  (-1)^{k(k-1)/2}
\sum_{\{\alpha_0\cdots \alpha_k\}} A^{(\alpha_0)} 
d A^{(\alpha_1)}\wedge \cdots \wedge dA^{(\alpha_k)} \nonumber 
\\ && \qquad  \qquad \wedge 
\bigg[\sum_{i=0}^k (-1)^{k-i} \chi_{n-k}^{\alpha_0\cdots \alpha_{i-1}\alpha_{i+1}
\cdots \alpha_k} + (-1)^{k+1}   
d \chi_{n-k-1}^{\alpha_0\cdots \alpha_k}\bigg] \, .
\een
We manipulate the second term inside the square bracket
by integration by parts and the first term by
noting that all $i$'s from 1 to $k$ gives identical contributions to the sum, so that
we can include the sum over $i=0$ and 1 only and multiply the result for $i=1$ by
a factor of $k$. After exchanging the labels $\alpha_0$ and $\alpha_1$ in the latter
term we get
\ben
\Delta &=& - \int_M \sum_{k=0}^K (-1)^{k(k+1)/2}
\sum_{\{\alpha_0\alpha_1\cdots \alpha_k\}}  (A^{(\alpha_0)} 
d A^{(\alpha_1)} - k\, A^{(\alpha_1)} dA^{(\alpha_0)}) \nonumber \\
&& \qquad \qquad \qquad \qquad \qquad \wedge dA^{(\alpha_2)}
\wedge \cdots \wedge dA^{(\alpha_k)}
\wedge \chi_{n-k}^{\alpha_1\cdots\alpha_k} \nonumber 
\\ && - \int_M \sum_{k=0}^K  (-1)^{k(k-1)/2}   
\sum_{\{\alpha_0\cdots \alpha_k\}} dA^{(\alpha_0)} \wedge
d A^{(\alpha_1)}\wedge \cdots \wedge dA^{(\alpha_k)} \wedge
\chi_{n-k-1}^{\alpha_0\cdots \alpha_k}\, .
\een
We can now perform the sum over $\alpha_0$ explicitly in the first term. Using
$\sum_\alpha A^{(\alpha)}(m)=1$ and $\sum_\alpha dA^{(\alpha)}=0$ we get
\ben
\Delta &=& - \int_M \sum_{k=1}^K (-1)^{k(k+1)/2}
\sum_{\{\alpha_1\cdots \alpha_k\}}  
d A^{(\alpha_1)} \wedge dA^{(\alpha_2)}
\wedge \cdots \wedge dA^{(\alpha_k)}
\wedge \chi_{n-k}^{\alpha_1\cdots\alpha_k} \nonumber 
\\ && - \int_M \sum_{k=0}^K (-1)^{k(k-1)/2} 
\sum_{\{\alpha_0\cdots \alpha_k\}} dA^{(\alpha_0)} \wedge
d A^{(\alpha_1)}\wedge \cdots \wedge dA^{(\alpha_k)} \wedge
\chi_{n-k-1}^{\alpha_0\cdots \alpha_k}\, .
\een
After renaming $k$ as $k+1$ in the first term we see that these two terms cancel, leading to
\be
\Delta = 0\, .
\ee

Next we turn to the proof of the second property. For this we note, using  the definition 
\refb{edefmunk} of $\mu$, (\ref{omox}),  the fact that $d\omega_n=0$, and the formula (\ref{ebdP}) for
$\partial P_{\alpha_0\cdots\alpha_k}$ that
\begin{align}\label{emuld}  
d\mu_{n-k}^{\alpha_0\cdots \alpha_k}&=d\int_{P_{\alpha_0\cdots \alpha_k}}\omega_n
= -(-1)^k 
\int_{\partial P_{\alpha_0\cdots\alpha_k}} \omega_n
= \sum_{i=0}^k(-1)^i\int_{P_{\alpha_0\cdots \alpha_{i-1}\alpha_{i+1}\cdots \alpha_k}}\omega_n\cr &=\sum_{i=0}^k(-1)^i\mu_{n-k+1}^{\alpha_0\cdots\alpha_{i-1}\alpha_{i+1}\dots \alpha_k}.
\end{align}

 Let us now consider an infinitesimal change\footnote{We can interpolate between any two partitions of unity $A^{(\alpha)}$ and $\t A^{(\alpha)}$ subordinate
 to the same open cover via the family of partitions of unity given by the functions $u A^{(\alpha)}+(1-u)\t A^{(\alpha)}$, $0\leq u\leq 1$. To show that the choice of
 partition of unity does not matter, it suffices to consider the effect of differentiating with respect to $u$.} 
 $A^{(\alpha)}\to A^{(\alpha)}+\delta A^{(\alpha)}$ subject to the constraint
 \refb{econsA}. This gives
 \be \label{econsdeltaA}
\delta A^{(\alpha)}(m) = 0 \quad \hbox{for $m\not\in \UU_\alpha$}, \quad 
\sum_\alpha \delta A^{(\alpha)}(m) = 0\, .
\ee
The change in the amplitude \refb{esmooth} 
under this infinitesimal change is given by
\ben \label{esmoothchange}
\delta \AAA &=& 
\int_M \sum_{k=0}^K (-1)^{k(k-1)/2}
 \sum_{\{\alpha_0,\cdots, \alpha_k\}} \left\{\delta A^{(\alpha_0)} 
d A^{(\alpha_1)} + k\, A^{(\alpha_0)} d \left(\delta A^{(\alpha_1)}\right) \right\}
\nonumber \\ && \qquad \qquad \qquad \qquad 
\wedge dA^{(\alpha_2)} \wedge \cdots \wedge dA^{(\alpha_k)} \wedge 
\mu_{n-k}^{\alpha_0\cdots\alpha_k}\, .
\een
We now manipulate the second term inside the curly bracket by 
integrating by parts
to move the $d$ operator from $\delta A^{(\alpha_1)}$  
to the rest of the terms and then
exchanging the 
labels $\alpha_0$ and $\alpha_1$, picking up a sign due to the antisymmetry of $\mu$.
This gives
\ben
\delta\AAA &=& \int_M \sum_{k=0}^K (-1)^{k(k-1)/2} \sum_{\{\alpha_0,\cdots, \alpha_k\}}
\delta A^{(\alpha_0)} 
d A^{(\alpha_1)} \wedge \cdots \wedge dA^{(\alpha_k)} \wedge 
\mu_{n-k}^{\alpha_0\cdots\alpha_k} \nonumber \\ &&
+ \int_M \sum_{k=0}^K (-1)^{k(k-1)/2} \sum_{\{\alpha_0,\cdots, \alpha_k\}} 
k \, \delta A^{(\alpha_0)} 
d A^{(\alpha_1)} \wedge \cdots \wedge dA^{(\alpha_k)} \wedge 
\mu_{n-k}^{\alpha_0\cdots\alpha_k} \nonumber \\ &&
+ \int_M \sum_{k=0}^K (-1)^{k(k-1)/2} \sum_{\{\alpha_0,\cdots, \alpha_k\}} 
k\, (-1)^{k-1} \, \delta A^{(\alpha_0)} 
A^{(\alpha_1)} \wedge dA^{(\alpha_2)} \nonumber \\ && \qquad  
 \qquad   \qquad   \qquad   \qquad   \qquad  
 \wedge \cdots \wedge dA^{(\alpha_k)} \wedge 
d\mu_{n-k}^{\alpha_0\cdots\alpha_k}\, .
\een
Combining the first two terms into a single term and replacing $d\mu$ by the right hand
side of \refb{emuld} in the last term we get
\ben
\delta\AAA &=& \int_M \sum_{k=0}^K (-1)^{k(k-1)/2} (k+1) 
\sum_{\{\alpha_0,\cdots, \alpha_k\}}
\delta A^{(\alpha_0)} 
d A^{(\alpha_1)} \wedge \cdots \wedge dA^{(\alpha_k)} \wedge 
\mu_{n-k}^{\alpha_0\cdots\alpha_k} \nonumber \\ &&
- \int_M \sum_{k=0}^K (-1)^{k(k-1)/2} \, k\, \sum_{\{\alpha_0,\cdots, \alpha_k\}} 
\delta A^{(\alpha_0)} 
A^{(\alpha_1)} \wedge dA^{(\alpha_2)}  \wedge \cdots \wedge dA^{(\alpha_k)} 
\nonumber \\ && \qquad  
 \qquad   \qquad   \qquad   \qquad   \qquad  
\wedge 
\sum_{i=0}^k (-1)^{k-i}  \mu_{n-k+1}^{\alpha_0\cdots\alpha_{i-1}\alpha_{i+1}\cdots
 \alpha_k}\, .
\een
We can now manipulate the second term by
noting that all $i$'s from 2 to $k$ gives identical contribution to the sum, so that
we can include the sum over $i=0,1$ and 2 only and multiply the result for $i=2$ by
a factor of $(k-1)$. This gives
\ben
\delta\AAA &=& \int_M \sum_{k=0}^K (-1)^{k(k-1)/2} (k+1) 
\sum_{\{\alpha_0,\cdots, \alpha_k\}}
\delta A^{(\alpha_0)} 
d A^{(\alpha_1)} \wedge \cdots \wedge dA^{(\alpha_k)} \wedge 
\mu_{n-k}^{\alpha_0\cdots\alpha_k} \nonumber \\ &&
- \int_M \sum_{k=0}^K (-1)^{k(k-1)/2} \, k\, \sum_{\{\alpha_0,\cdots, \alpha_k\}} 
\delta A^{(\alpha_0)} 
A^{(\alpha_1)} \wedge dA^{(\alpha_2)}  \wedge \cdots \wedge dA^{(\alpha_k)} 
\nonumber \\ && \qquad  
\wedge  \left\{ (-1)^k \mu_{n-k+1}^{\alpha_1\cdots\alpha_k} 
+ (-1)^{k-1} \mu_{n-k+1}^{\alpha_0\alpha_2\cdots\alpha_k} 
+ (-1)^{k-2} (k-1) \mu_{n-k+1}^{\alpha_0\alpha_1\alpha_3\cdots\alpha_k} 
 \right\}
 \, . \nonumber \\
\een
The contribution from the first term in the curly bracket vanishes 
since $\sum_{\alpha_0}\delta A^{(\alpha_0)}=0$. The second term can be simplified
using $\sum_{\alpha_1} A^{(\alpha_1)}=1$. The third term vanishes since
$\sum_{\alpha_2} dA^{(\alpha_2)}=0$.  This gives
\ben  
\delta\AAA &=& \int_M \sum_{k=0}^K (-1)^{k(k-1)/2} (k+1) 
\sum_{\{\alpha_0,\cdots, \alpha_k\}}
\delta A^{(\alpha_0)} 
d A^{(\alpha_1)} \wedge \cdots \wedge dA^{(\alpha_k)} \wedge 
\mu_{n-k}^{\alpha_0\cdots\alpha_k} \nonumber \\ &&
- \int_M \sum_{k=1}^K \, k\, (-1)^{(k-1)(k-2)/2} \sum_{\{\alpha_0,\alpha_2,\cdots, \alpha_k\}} 
\delta A^{(\alpha_0)} 
\wedge dA^{(\alpha_2)}  \wedge \cdots \wedge dA^{(\alpha_k)} 
\wedge \mu_{n-k+1}^{\alpha_0\alpha_2\cdots\alpha_k} 
 \, . \nonumber \\
\een
Relabelling $k$ as $k+1$ in the second term we see that the two terms cancel,
leading to
\be
\delta \AAA=0\, .
\ee

The third 
property -- that  
the amplitude does not depend on the choice of sections $s^\alpha$ -- is
an immediate consequence.
Let us suppose that
we want to change the section on a particular $\UU_\alpha$ -- call it $\UU_*$ --
from $s^{*_1}$ to $s^{*_2}$, satisfying
\be \label{ereq*}
(m;s^{*_i}(m), s^{\alpha_1}, 
\cdots s^{\alpha_k}(m))\in X^{(k+1)} \quad 
\hbox{for $m\in \UU_{*}\cap \UU_{\alpha_1} \cdots \cap \UU_{\alpha_k}$}, \quad
i=1,2\, .
\ee
An alternative representation of these two choices of $s^*$ can be given as follows.
Let us consider a cover of $M$ that is identical to the original choice except that
the open set $\UU_*$ occurs twice. Call the two copies $\UU_{*_1}$ and $\UU_{*_2}$, and
choose sections $s^{*_1}$ and $s^{*_2}$ on them. Then the choice $s^*=s^{*_1}$
on $\UU_*$
will correspond to choosing $A^{(*_1)}=A^{(*)}$, $A^{(*_2)}=0$ and 
the choice $s^*=s^{*_2}$ on $\UU_*$
will correspond to choosing $A^{(*_2)}=A^{(*)}$, $A^{(*_1)}=0$.  Since the choice of a partition of unity does not matter, the  two
choices $s^{*_1}$ and $s^{*_2}$ of $s^*$ give identical results for the amplitude.

An attentive reader might notice that we have skipped over a fine point here.
Since $\UU_{*_1}$ has complete overlap with $\UU_{*_2}$, the above
analysis requires that
\be \label{ereqdnew}
(m;s^{*_1}(m), s^{*_2}(m), s^{\alpha_1}(m), \cdots s^{\alpha_k}(m))\in X^{(k+2)} \quad 
\hbox{for $m\in \UU_{*}\cap \UU_{\alpha_1} \cdots \cap \UU_{\alpha_k}$}\, .
\ee
This is somewhat stronger than \refb{ereq*} and can fail in a non-generic situation
even if \refb{ereq*} holds. However we can circumvent this problem by choosing a
third section $s^{\times}$ on $\UU_*$ satisfying
\be \label{ereqdthird}
(m;s^{*_i}(m), s^{\times}(m), s^{\alpha_1}(m), \cdots s^{\alpha_k}(m))\in X^{(k+2)} 
\quad \hbox{for $m\in \UU_{*}\cap \UU_{\alpha_1} \cdots \cap \UU_{\alpha_k}$}
\quad
i=1,2\, .
\ee
The existence of $s^{(\times)}$ satisfying \refb{ereqdthird} can be proved using the
method of section \ref{ssection} for a sufficiently fine covering. Now our previous argument
can be used to show that the result for the amplitude for 
the sections $s^{*_1}$ and $s^{*_2}$ are identical to that for
section $s^\times$, and hence the results for the choices 
$s^{*_1}$ and $s^{*_2}$ are identical to  each other.

The next property -- that the amplitude does not depend on the choice of an open cover -- 
also follows from the second property.  
Let $\S$ be an open cover by open sets $\UU_\alpha,$ $\alpha\in I$, and $\S'$ another
open cover by open sets $\VV_\beta,\,\beta\in J$.  We can define a third open cover $\S''$ in which the open
sets $\WW_\sigma$ are labeled by the union $I\cup J$, with $\WW_\sigma=\UU_\sigma$ for $\sigma\in I$
and $\WW_\sigma=\VV_\sigma$ for $\sigma\in J$.  In defining the amplitude using the open cover $\S''$, we have a lot
of freedom in the choice of a partition of unity.  We can pick the partition of unity such that $A^\sigma=0$ for $\sigma\in J$.  This implies that the $A^\sigma$, $\sigma\in I$,
are a partition of unity subordinate to the original open cover $\S$.
In this case, the amplitude computed using the open cover $\S''$ immediately reduces to what we would have gotten
using the open cover $\S$.  Alternatively, reversing the roles of the finite sets $I$ and $J$, we could pick a partition of unity subordinate to $\S''$ such that the calculation
of the amplitude reduces to what we would have gotten using the cover $\S'$.  Hence  any choice of open cover leads to the same amplitudes. 

Finally, we want to show that the amplitude 
\refb{esmooth} computed via a general open cover and partition of unity coincides with the amplitude \refb{efull} computed using a dual triangulation.
We shall do this by showing
that
given the data used in section \ref{amplitude}, i.e. the dual 
triangulation and the choice of
local section on each polyhedron, we can choose  a covering by open sets $\UU_\alpha$, a
partition of unity $\{A^{(\alpha)}\}$, and local sections $s^\alpha$ so that the formula \refb{esmooth}  gives us back the result of section \ref{amplitude}. This is
done as 
follows:
\begin{enumerate}
\item First we shall describe the choice of the open sets.
Given any polyhedron $M_0^\alpha$ forming part of a dual triangulation, 
we thicken it slightly to make an  open set $\UU_{\alpha}$.  This gives an open cover of $M$, with  the property that  $\UU_{\alpha}\cap \UU_\beta$ 
is a slight thickening $M_0^\alpha\cap M_0^\beta$, and similarly for multiple intersections.
\item Next we describe the choice of the local sections
$s^\alpha$ on each $\UU_{\alpha}$ and the partition of
unity $A^{(\alpha)}$.
We  choose the local
sections $s^{\alpha}:\UU_\alpha\to X$ 
so that their restrictions to $M_0^\alpha$ are the sections $s^\alpha:M_0^\alpha\to X$ 
that were used in section \ref{amplitude}.
Furthermore, we
choose $A^{(\alpha)}$ to be a slightly smoothed
version of the characteristic function of $M_0^\alpha$ (the function that is 
 1 inside $M_0^\alpha$ and  0 outside), which we will call $H_\alpha$.  \end{enumerate}

With this choice, the amplitude \refb{esmooth} is given by
\be \label{einter}
\AAA=\int_M \sum_{k=0}^K (-1)^{k(k+1)/2} \,
(-1)^k \sum_{\{\alpha_0,\cdots, \alpha_k\}} A^{\alpha_0}
d A^{\alpha_1}\wedge \cdots \wedge d A^{\alpha_k} \wedge 
\mu_{n-k}^{\alpha_0\cdots\alpha_k}\, ,
\ee
where $A^\alpha$ is a slightly  
smoothed version of the characteristic function $H_\alpha$.
Let $\PP_{\alpha_0\cdots\alpha_k}$ denote the operation of 
summing over all permutations $P$ of $\alpha_0,\cdots \alpha_k$ 
weighted by $(-1)^P$.
We shall show that
in the limit $A^\alpha\to H^\alpha$, 
\be \label{edeltareg}
\rho_k^{\alpha_0\cdots\alpha_k}
\equiv (-1)^k \PP_{\alpha_0\cdots\alpha_k} [A^{\alpha_0} 
d A^{\alpha_1}\wedge \cdots \wedge d A^{\alpha_k}]
\ee
approaches the 
$\delta$-function that localizes the integral on the subspace 
$M_k^{\alpha_0\cdots\alpha_k}$. Thus we get back \refb{efull}.

This result 
together with the previous results of this section immediately shows that the
amplitude \refb{efull} is independent of the choice of dual triangulation and the
choice of the sections $\{s^\alpha\}$ on the codimension zero faces $\{M_0^\alpha\}$
used in the construction of section \ref{amplitude}. 

It remains to prove that, in the limit $A^\alpha\to  H^\alpha$, the right hand side of  \refb{edeltareg} 
approaches a delta function supported on
$M_k^{\alpha_0\cdots\alpha_k}$.  In this limit, each factor $d A^{\alpha_i}$ converges to a delta function
with support in codimension 1.  Each term on the right hand side of eqn. (\ref{edeltareg}) is a product of $k$ such
terms and so will have delta function support in codimension $k$.  
It is clear that this support is localized on 
$M_k^{\alpha_0\cdots\alpha_k}$, since the $d A^{\alpha_i}$ factor vanishes
outside $M_0^{\alpha_i}$ and the $A^{\alpha_0}$ factor vanishes outside
$M_0^{\alpha_0}$. Thus we only have to show that the normalization
is correct. We shall prove this inductively, i.e. assuming that it holds up to
a certain value of $k$, we shall prove that it holds when we increase $k$
by 1. Since this is manifestly true for $k=0$ -- a codimension zero delta function
with support on $M_0^{\alpha_0}$ being simply the characteristic function of $M_0^{\alpha_0}$ --
the result follows for general $k$.

We  take a small 
tubular neighbourhood $T^{\alpha_0\cdots \alpha_k}$ of
$M_k^{\alpha_0\cdots \alpha_k}$ and foliate it 
by a family of dimension $k$ balls $B_k(m)$, intersecting 
$M_k^{\alpha_0\cdots \alpha_k}$ transversely at the point
$m\in M_k^{\alpha_0\cdots \alpha_k}$.
We take the size of the ball $B_k$ 
to be 
large compared to the regulator used to approximate $H^\alpha$ by
$A^\alpha$, but sufficiently small so as not to intersect any 
$M_0^\alpha$ other than $M_0^{\alpha_0},\cdots M_0^{\alpha_k}$.
The orientation of $B_k$ is chosen so that locally 
$B_k\times M_k^{\alpha_0\cdots \alpha_k}$ has the same orientation as 
$M$. 
To prove the desired result,
we need to show that the integral of
 $\rho_k^{\alpha_0\cdots\alpha_k}$ over $B_k$ gives 1. Now 
inside $B_k$, all $A^\alpha$'s vanish except for $\alpha=\alpha_0,\cdots \alpha_k$
 and hence we have 
\be
A^{\alpha_0}=1 - \sum_{j=1}^k A^{\alpha_j}, \quad
 dA^{\alpha_0}= - \sum_{j=1}^k dA^{\alpha_k}\, .
 \ee
Using this, we can express $\rho_k$ given in \refb{edeltareg} as
\ben
&& \rho_k^{\alpha_0\cdots\alpha_k}\nonumber \\
&=&  (-1)^k \, k!\, \Big[
A^{\alpha_0} 
d A^{\alpha_1}\wedge \cdots \wedge d A^{\alpha_k}
- \sum_{i=1}^k A^{\alpha_i} 
d A^{\alpha_1}\wedge \cdots dA^{\alpha_{i-1}} \wedge dA^{\alpha_0}\wedge
dA^{\alpha_{i+1}} \cdots \wedge d A^{\alpha_k}\Big] \nonumber \\
&=& (-1)^k \, k!\, \Big[
\Big( 1 - \sum_{i=1}^k A^{\alpha_i}\Big)
d A^{\alpha_1}\wedge \cdots \wedge d A^{\alpha_k} \nonumber \\ &&
+ \sum_{i=1}^k A^{\alpha_i} 
d A^{\alpha_1}\wedge \cdots dA^{\alpha_{i-1}} \wedge dA^{\alpha_i}\wedge
dA^{\alpha_{i+1}} \cdots \wedge d A^{\alpha_k}\Big] \nonumber \\
&=& (-1)^k \, k!\, d A^{\alpha_1}\wedge \cdots \wedge d A^{\alpha_k} 
= - d \rho_{k-1}^{\alpha_1\cdots \alpha_{k-1}}\, .
\een
This gives
\be \label{eint1}
\int_{B_k}\rho_k^{\alpha_0\cdots\alpha_k} = -\int_{\p B_k} 
\rho_{k-1}^{\alpha_1\cdots\alpha_k}\, .
\ee
Our earlier arguments show that $\rho_{k-1}^{\alpha_1\cdots\alpha_k}$
has support only in the neighbourhood of $M_{k-1}^{\alpha_1\cdots
\alpha_k}$. $\p B_{k}$ intersects it at some point $m'$. Let 
$B_{k-1}(m)$ for $m\in M_{k-1}^{\alpha_1\cdots \alpha_k}$ 
denote a family of $(k-1)$-dimensional balls 
centered at $m$ that can be used to foliate a tubular neighbourhood of 
$T^{\alpha_1\cdots
\alpha_k}$ of $M_{k-1}^{\alpha_1\cdots
\alpha_k}$. We pick the orientation of $B_{k-1}$
such that locally $B_{k-1}\times M_{k-1}^{\alpha_1\cdots
\alpha_k}$ has the same orientation as $M$. 
Then the relevant part of $\p B_k(m)$ in \refb{eint1} can be
replaced by $\pm B_{k-1}(m')$ with the sign determined by comparing the
orientations of $\p B_k$ and $B_{k-1}$. We shall soon show that the sign
is negative. This gives
\be 
\int_{B_k(m)}\rho_k^{\alpha_0\cdots\alpha_k} =
\int_{B_{k-1}(m')} \rho_{k-1}^{\alpha_1\cdots\alpha_k}\, .
\ee
But  we have assumed that $\rho_{k-1}^{\alpha_1\cdots\alpha_k}$
gives the delta function that localizes the integral on 
$M_{k-1}^{\alpha_1\cdots\alpha_k}$. Thus the right hand side is 1 and
we get the desired result.

Let us now show that $\p B_k$ and $B_{k-1}$ have opposite orientation.
For this we note that
\be \label{eb1}
\p T^{\alpha_0\cdots \alpha_k}
= \p B_k \times M_k^{\alpha_0\cdots \alpha_k}
+ (-1)^k B_k\times \p M_k^{\alpha_0\cdots \alpha_k}\, ,
\ee
where we have used $\times$ to denote fibering, {\it e.g.} the first term
on the right hand side denotes $\p B_k$ fibered over 
$M_k^{\alpha_0\cdots \alpha_k}$.
On the other hand we have
\be
\p T^{\alpha_1\cdots \alpha_k}
= \p B_{k-1}\times M_{k-1}^{\alpha_1\cdots \alpha_k} 
+ (-1)^{k-1} B_{k-1}\times \p M_{k-1}^{\alpha_1\cdots \alpha_k}
\ee
Using \refb{eorientation} we see that the second term on the right hand side has a component
\be
- (-1)^{k-1} B_{k-1}\times M_k^{\alpha_1\cdots \alpha_k\alpha_0}
= B_{k-1}\times M_k^{\alpha_0\cdots\alpha_k}\, .
\ee
This must be oppositely oriented to the component 
$\p B_k \times M_k^{\alpha_0\cdots \alpha_k}$ in \refb{eb1} since
$T^{\alpha_0\cdots \alpha_k}$ and $T^{\alpha_1\cdots \alpha_k}$
are complementary subspaces of $M$.
Thus we see that $\p B_k$ has opposite orientation to $B_{k-1}$.

\section{Decoupling of Pure Gauge States} \label{decoupling}  

It remains to establish gauge invariance, which states that the amplitude vanishes if
 all  external states are BRST-invariant, and one of them is also BRST-trivial.  
It is well known (see footnote \ref{fo4}) 
that in this case
\be \label{eomlamb}
\omega_n=d\lambda_{n-1}
\ee 
where
$\lambda_{n-1}$ has a form similar to that in \refb{edefomega}
\be \label{edeflambda}
\lambda_{n-1} = \left\langle \prod_{i=1}^K (\XX(z_i) - \p\xi(z_i) dz^i) \wedge \OO'
\right\rangle_{n-1}\, .
\ee 
We shall carry out our analysis for the amplitude
\refb{esmooth} since \refb{efull} can be regarded as a special case.
Integrating \refb{eomlamb} over 
$P_{\alpha_0\cdots\alpha_k}$ for fixed $m\in \UU_{\alpha_0}\cup
\cdots \UU_{\alpha_k}$, and using \refb{omox}
we get
\be
\mu_{n-k}^{\alpha_0\cdots\alpha_k} 
= \int_{P_{\alpha_0\cdots\alpha_k}} \omega_n =  \int_{P_{\alpha_0\cdots\alpha_k}} d
\lambda_{n-1} =
\left[ (-1)^k 
d \int_{P_{\alpha_0\cdots\alpha_k}} \lambda_{n-1}
+ \int_{\p P_{\alpha_0\cdots\alpha_k}} \lambda_{n-1}\right]\, .
\ee
Defining 
\be
\nu_{n-k-1}^{\alpha_0\cdots \alpha_k} = 
\int_{P_{\alpha_0\cdots\alpha_k}} \lambda_{n-1}\, ,
\ee
and using \refb{ebdP} we get the relation
\be \label{emunu}
\mu_{n-k}^{\alpha_0\cdots\alpha_k}  =  \left[ (-1)^k
d\nu_{n-k-1}^{\alpha_0\cdots \alpha_k} - \sum_{i=0}^k (-1)^{k-i}
\nu_{n-k}^{\alpha_0\cdots\alpha_{i-1}\alpha_{i+1}\cdots \alpha_k}\right]\, .
\ee 

Substituting \refb{emunu} into \refb{esmooth} we now get
\ben
\AAA &=&
  \, \int_M \sum_{k=0}^K (-1)^{k(k-1)/2}
 \sum_{\{\alpha_0,\cdots, \alpha_k\}} A^{(\alpha_0)} 
d A^{(\alpha_1)}\wedge \cdots \wedge dA^{(\alpha_k)} \nonumber \\
&& \qquad  \qquad  \qquad  \qquad \wedge 
\left[ (-1)^k
d\nu_{n-k-1}^{\alpha_0\cdots \alpha_k} - \sum_{i=0}^k (-1)^{k-i}
\nu_{n-k}^{\alpha_0\cdots\alpha_{i-1}\alpha_{i+1}\cdots \alpha_k}\right]\, .
\een
We analyze the first term by integration by parts. For the second term we note
that all $i$ from 1 to $k$ give identical results; so we can just restrict the sum to $i=0$
and $i=1$ and multiply the latter by a factor of $k$. This gives
\ben
\AAA &=& -\int_M \sum_{k=0}^K (-1)^{k(k-1)/2} 
 \sum_{\{\alpha_0,\cdots, \alpha_k\}} 
d A^{(\alpha_0)} \wedge
d A^{(\alpha_1)}\wedge \cdots \wedge dA^{(\alpha_k)} \wedge 
\nu_{n-k-1}^{\alpha_0\cdots \alpha_k}  \nonumber \\
&& - \int_M \sum_{k=0}^K (-1)^{k(k-1)/2} \sum_{\{\alpha_0,\cdots, \alpha_k\}}  
A^{(\alpha_0)} \wedge
d A^{(\alpha_1)}\wedge \cdots \wedge dA^{(\alpha_k)} 
\nonumber \\ && \qquad   \qquad  \qquad  \qquad \wedge
\left[(-1)^k \nu_{n-k}^{\alpha_1\cdots \alpha_k} +
(-1)^{k-1} \, k\,  \nu_{n-k}^{\alpha_0\alpha_2\alpha_3\cdots \alpha_k}\right]\, .
\een
For the first term inside the square bracket in the last line we can perform the sum
over $\alpha_0$ using $\sum_{\alpha_0}A^{(\alpha_0)}=1$ and for the second
term inside the square bracket in the last line we can perform the sum
over $\alpha_1$ using $\sum_{\alpha_1}dA^{(\alpha_1)}=0$. This gives
\ben
\AAA &=& -\int_M \sum_{k=0}^K  (-1)^{k(k-1)/2} \sum_{\{\alpha_0,\cdots, \alpha_k\}} 
d A^{(\alpha_0)} \wedge
d A^{(\alpha_1)}\wedge \cdots \wedge dA^{(\alpha_k)} \wedge 
\nu_{n-k-1}^{\alpha_0\cdots \alpha_k}  \nonumber \\
&& - \int_M \sum_{k=1}^K(-1)^{k(k+1)/2}  \sum_{\{\alpha_1,\cdots, \alpha_k\}}  
d A^{(\alpha_1)}\wedge \cdots \wedge dA^{(\alpha_k)} \wedge
\nu_{n-k}^{\alpha_1\cdots \alpha_k} \, .
\een
Replacing $k$ by $k+1$ in the last term and using the fact that $(-1)^{(k+1)(k+2)/2}
= -(-1)^{k(k-1)/2}$ we see that the two terms cancel.

\section{Interpretation Via Super Riemann Surfaces}\label{super}

We conclude by describing how one would interpret these results from the point of view of super Riemann
surface theory.

A perturbative scattering amplitude in superstring theory is naturally understood as the integral of a natural measure $\Psi$ on, roughly
speaking, the moduli space $\MM$ of super Riemann surfaces.\footnote{\label{elide} We elide some details that are most fully explained in section 5
of \cite{WittenNotes}.  To be more precise, instead of $\MM$ one should consider an appropriate cycle $\varGamma\subset \MM_\ell\times \MM_r$,
where $\MM_r$ and $\MM_\ell$ parametrize respectively the holomorphic and antiholomorphic complex structure of the superstring worldsheet.
For example, for the heterotic string, $\MM_r$ is the moduli space $\MM$ of super Riemann surfaces, $\MM_\ell$ is the analogous bosonic moduli space
$M$ (with its complex structure reversed) and one can think of $\varGamma$ as $\MM$ with a choice of smooth structure.}  
The worldsheet path integral determines a natural measure on $\MM$, and perturbative superstring scattering amplitudes are obtained by integrating this
measure.

What do PCO's mean in this framework?  This question was answered long ago \cite{VV}.  PCO's are a method to parametrize the odd directions
in $\MM$ by the use of $\delta$-function gravitino perturbations.\footnote{PCO's have an analog in the theory of ordinary Riemann surfaces in the form of Schiffer variations.  A Schiffer variation
is a deformation of the complex structure of a Riemann surface $\Sigma$ that is determined by a change in the metric of $\Sigma$ with $\delta$-function
support.  (The phrase ``Schiffer variation'' is also sometimes used to refer to a deformation  with support in a very small open set.)} Any local choice of PCO's (avoiding spurious singularities) gives a parametrization
of an open set in $\MM$, and gives a valid way to compute the superstring measure $\Psi$ in that open set.  

From this point of view, then, the way to use PCO's is to cover $M$ with open sets $\UU_\alpha$, and pick in each $\UU_\alpha$ a section $s^\alpha:\UU_\alpha
\to X$ corresponding to a local choice of PCO's.  This will give a convenient way to calculate
 a superstring measure $\Psi_\alpha$ on an open set $\U_\alpha\subset \MM$
that corresponds to $\UU_\alpha\subset M$.  If $\UU_\alpha$ and $\UU_\beta$ are open sets in $M$ with local sections $s^\alpha$ and $s^\beta$,
then $\Psi_\alpha$ and $\Psi_\beta$ are equal on $\U_\alpha\cap \U_\beta$.  (We stress that they are equal, not equal up to a total derivative.)
Thus the local measures $\Psi_\alpha$ computed using PCO's on  the open sets $\U_\alpha$ making up a cover of $\MM$
automatically glue together to determine the superstring measure $\Psi$ on $\MM$. More on this can be found in  sections 3.5-6 of \cite{Revisited}.
The perturbative
superstring scattering amplitude is simply defined as $\int_\MM\Psi$.    Gauge-invariance follows from the super-analog of Stoke's theorem, which
says that if $\Psi=d\Lambda$ (where now $\Lambda$ is an integral form of codimension 1, a concept described for example in \cite{WittenNotes}), then
$\int_\MM\Psi=\int_\MM d\Lambda=0$.

As for explicitly computing an integral $\int_\MM\Psi$, there are various possible approaches.  For a smooth supermanifold $\MM$ (see footnote \ref{elide}) with
reduced space $M$,
one can always pick a smooth projection $\zeta:\MM\to M$.  By integrating over the fibers of $\zeta$, one gets a smooth measure $\zeta_*(\Psi)$ on $M$
and
\be\label{torf} \int_\MM\Psi=\int_M\zeta_*(\Psi).  \ee
Thus for any choice of the smooth projection $\zeta$, the superstring scattering amplitude can be computed by integration over $M$ of a smooth
measure, namely $\zeta_*(\Psi)$.  However, while the underlying measure $\Psi$ on $\MM$ is completely natural, the choice of $\zeta$ is not and
the induced measure
$\zeta_*(\Psi)$ on $M$ depends on that choice.  If $\zeta$ and $\t\zeta$ are two smooth projections from $\MM$ to $M$,
then $\zeta_*(\Psi)-\t\zeta_*(\Psi)=d\chi$, for some $(n-1)$-form $\chi$ on $M$.  Hence $\int_M\zeta_*(\Psi)=\int_M\t\zeta_*(M)$, in keeping with the
fact that they both equal $\int_\MM\Psi$.
All this is in accord with what we found in section \ref{smooth}: superstring amplitudes can be computed by integrating a smooth measure on
the bosonic moduli space $M$, but there is no natural choice of this smooth measure. 

Sometimes it may be inconvenient to pick a globally-defined smooth projection $\zeta:\MM\to M$, or we may not wish to do so.  Then we can
proceed as follows.  We pick a dual triangulation $\Upsilon$ of $M$ (or some more general covering).  In a small neighborhood of
each polyhedron $M_0^\alpha$, we pick a smooth splitting $\zeta^\alpha: \MM\to M$.  We do not impose any compatibility 
between the different $\zeta^\alpha$.  We cannot,
therefore, expect a simple relation
\be\label{zonk}\int_\MM\Psi\overset{?}= \sum_\alpha \int_{M_0^\alpha}\zeta^\alpha_*(\Psi). \ee
To compensate for the mismatch between $\zeta^\alpha$ and $\zeta^\beta$ along $M_0^\alpha\cap M_0^\beta=M_1^{\alpha\beta}$, one must add a correction term along $M_1^{\alpha\beta}$.
There is no unique way to determine this correction term.  For example (though this is certainly not the only approach), 
one might correct $\zeta^\alpha$ and $\zeta^\beta$ very near $M_1^{\alpha\beta}$
so that they agree; any way to do this will give a correction term along $M_1^{\alpha\beta}$ that should be added to the right hand side of eqn. (\ref{zonk}).
If this is done properly -- but this may be inconvenient in practice -- 
 the corrected $\zeta^\alpha$, $\zeta^\beta$, and $\zeta^\gamma$ will agree along the triple intersections $M_2^{\alpha\beta\gamma}$.
Otherwise further corrections supported on $M_2^{\alpha\beta\gamma}$ must be added.  Again, if those corrections are made independently, without worrying
about compatibility on quadruple overlaps, then one will require further corrections on $M_3^{\alpha\beta\gamma\delta}$.  In general, 
one will go on in this way all the
way down to codimension $n$, at which point the process will stop and we will get a formula for $\int_\MM\Psi$ as 
a sum of contributions from the polyhedra in $\Upsilon$
and their faces of various codimension.
The general structure is very similar to what we found in section \ref{amplitude}, and in fact what was just explained was part of  the motivation for that
construction.  (The construction in section \ref{amplitude} terminated in codimension $K<n$, where $K$ is the odd dimension of $\M$.  This is related
to the fact that the expansion of a function on $\M$ in powers of the odd coordinates terminates with the $K$-th term, which makes it natural
for a construction along the lines just explained to terminate in codimension $K$.)

Alternatively, we can compute using a partition of unity.  One approach is as follows.  We pick a  cover of $\M$ by open sets $\U_\alpha$ (which are in
1-1 correspondence with open sets $\UU_\alpha\subset M$).  We pick a partition of unity on $\M$ subordinate to the cover by the $\U_\alpha$; as on a bosonic manifold,
this means that we pick functions $A^{(\alpha)}$ on $\MM$ that vanish outside $\U_\alpha$ and obey $\sum_\alpha A^{(\alpha)}=1$.  (Technically, it is convenient to require also
that the closure in $\M$ of the support of $A^{(\alpha)}$ should be contained in $\U_\alpha$.)  The $A^{(\alpha)}$ can be restricted to $M$
to give a partition of unity on $M$ (subordinate to the open cover by the $\UU_\alpha$), but the partition of unity on $\M$ carries in a certain sense more information, as we will see.  Since $\sum_\alpha A^{(\alpha)}=1$, and $A^{(\alpha)}$
vanishes outside $\U_\alpha$,
we have trivially
\begin{equation}\label{dorf}\int_\M\Psi=\sum_\alpha\int_\M A^{(\alpha)}\Psi=\sum_\alpha \int_{\U_\alpha}A^{(\alpha)} \Psi. \end{equation}
Given this, a partition of unity can be used to correct the naive equation (\ref{zonk}).  If $s^\alpha:\U_\alpha\to \UU_\alpha\subset M$ is any projection, then
\begin{equation}\label{morf}\int_{\U_\alpha}A^{(\alpha)}\Psi=\int_{\UU_\alpha}s^\alpha_*(A^{(\alpha)}\Psi).  \end{equation}
So  a perturbative superstring scattering amplitude can be calculated using a partition of unity as follows:
\begin{equation}\label{zorf}\int_\M\Psi=\sum_\alpha\int_{\UU_\alpha}s^\alpha_*(A^{(\alpha)}\Psi). \end{equation}
There is no need here for any compatibility between $s^\alpha$ and $s^\beta$ on $\U_\alpha\cap \U_\beta$.  Since $s^\alpha_*(A^{(\alpha)} \Psi)$ vanishes outside $\UU_\alpha$,
another equivalent formula is
\begin{equation}\label{worf}\int_\M\Psi=\int_M\sum_\alpha s^\alpha_*(A^{(\alpha)}\Psi). \end{equation}

For another way to use a partition of unity on $\M$, we recall that, as explained in textbooks, such a partition of unity can be used to construct a smooth projection  $\zeta:\M\to M$, giving one route to a formula along the lines of eqn. (\ref{torf}).

The reader may be curious to understand more explicitly the difference between the correct formula (\ref{dorf}) and the wrong formula (\ref{zonk}).  Let $A^{(\alpha)}_0$ be a partition of unity on $M$
subordinate to a cover by open sets $\UU_\alpha$.  This means in particular that $\sum_\alpha A^{(\alpha)}_0=1$.  
We can pull back the $A^{(\alpha)}_0$ to functions $(s^\alpha)^*(A^{(\alpha)}_0)$ on $\M$.  However, these functions do not give a partition
of unity on $\M$, since $\sum_\alpha (s^\alpha)^*(A^{(\alpha)}_0)$ equals 1 on $M$ but not necessarily on $\M$.  By adding nilpotent terms, the functions  $(s^\alpha)^*(A^{(\alpha)}_0)$ can
be corrected to functions $A^{(\alpha)}$ that give a partition of unity on $\M$.  Eqn. (\ref{zonk}) is analogous to trying to use  $(s^\alpha)^*(A^{(\alpha)}_0)$ rather than $A^{(\alpha)}$ in 
(\ref{dorf}).

We conclude with the following remark.   There are two possible points of view on superstring perturbation theory.
One may  view PCO's, supplemented with some procedure
such as the one described in the present paper, as the basic definition.
Or one may view the basic definition as being provided by integration over moduli of super Riemann surfaces,
with PCO's regarded as an often convenient method of computation. 
\vskip 1cm
\noindent
{\bf{Acknowledgments}}
AS thanks R. Pius and A. Rudra for discussions.  Research of AS was supported in part by the 
DAE project 12-R\&D-HRI-5.02-0303 and the J. C. Bose fellowship of 
the Department of Science and Technology, India.
EW thanks R. Donagi for discussions. 
Research of EW was supported in part by NSF Grant PHY-1314311. 

\bibliographystyle{unsrt}

\begin{thebibliography}{99}
\bibitem{FMS}
D. Friedan, E. Martinec, and S. Shenker, ``Covariant Quantization Of Superstrings,'' Phys. Lett. {\bf B160} (1985) 55,
``Conformal Invariance, Supersymmetry, and String Theory,'' Nucl. Phys. {\bf B271} (1986) 93.

\bibitem{VV}
E. Verlinde and H. Verlinde, ``Multiloop Calculations In Covariant Superstring Theory,'' Phys. Lett. {\bf B192} (1987) 95. 


\bibitem{Sen}
 A.~Sen,
  ``Off-shell Amplitudes in Superstring Theory,''
  arXiv:1408.0571v3 [hep-th].


\bibitem{WittenNotes}
E. Witten, ``Notes On Supermanifolds And Integration,'' arXiv:1209.2199.

\bibitem{Revisited}
E. Witten, ``Superstring Perturbation Theory Revisited,'' arXiv:1209.5461.

\bibitem{EKS}
T. Erler, S. Konopka, and I. Sachs, ``Resolving Witten's Superstring Field Theory,''  JHEP 04 (2014) 150,
arXiv:1312.2948.


\end{thebibliography}

\end{document}